\documentclass[12pt,twoside]{article}
\usepackage{psfig}
\newcommand{\VolumeHeader}{}
\newcommand{\VolumeSerial}{LNS}

\newcommand{\ActivityName}{ {\normalsize {\it 
Summer School on Astroparticle Physics and Cosmology
}}}
\newcommand{\ActivityDate}{ {\normalsize {\it
Trieste, 12-30 June 2000
}}}

\def\la{\mathrel{\mathpalette\fun <}}
\def\ga{\mathrel{\mathpalette\fun >}}
\def\fun#1#2{\lower3.6pt\vbox{\baselineskip0pt\lineskip.9pt
  \ialign{$\mathsurround=0pt#1\hfil##\hfil$\crcr#2\crcr\sim\crcr}}}

\newcommand{\LectureHeader}{Ultrahigh Energy Cosmic Rays...}

\begin{document}
\pagestyle{myheadings}
\markboth{\LectureHeader}{\VolumeHeader}
\markright{\VolumeHeader}


\begin{titlepage}


\title{Particle and Astrophysics Aspects of Ultrahigh Energy Cosmic Rays}

\author{G. Sigl$^\dagger$\thanks{sigl@iap.fr}
\\[1cm]
{\normalsize
{\it $^\dagger$ Institut d'Astrophysique de Paris, Paris, France}}
\\[10cm]
{\normalsize {\it Lecture given at the: }}
\\
\ActivityName 
\\
\ActivityDate 
\\[1cm]
{\small \VolumeSerial} 
}
\date{}
\maketitle
\thispagestyle{empty}
\end{titlepage}

\baselineskip=14pt
\newpage
\thispagestyle{empty}


\begin{abstract}
The origin of cosmic rays is one of the major unresolved astrophysical
questions. In particular,
the highest energy cosmic rays observed possess macroscopic energies and
their origin is likely to be associated with the most energetic processes
in the Universe. Their existence triggered a flurry of theoretical
explanations ranging from conventional shock acceleration to particle
physics beyond the Standard Model and processes taking place at the
earliest moments of our Universe. Furthermore, many new experimental
activities promise a strong increase of statistics at the highest
energies and a combination with $\gamma-$ray and neutrino astrophysics
will put strong constraints on these theoretical models. Detailed Monte
Carlo simulations indicate that charged ultra-high energy cosmic rays
can also be used as probes of large scale magnetic fields whose origin
may open another window into the very early Universe. We give an overview
over this quickly evolving research field.
\end{abstract}

\vspace{6cm}


{\it PACS numbers:}


\newpage
\thispagestyle{empty}
\tableofcontents

\newpage
\setcounter{page}{1}


\section{Introduction}

\begin{figure}[htb]
\centerline{\hbox{\psfig{figure=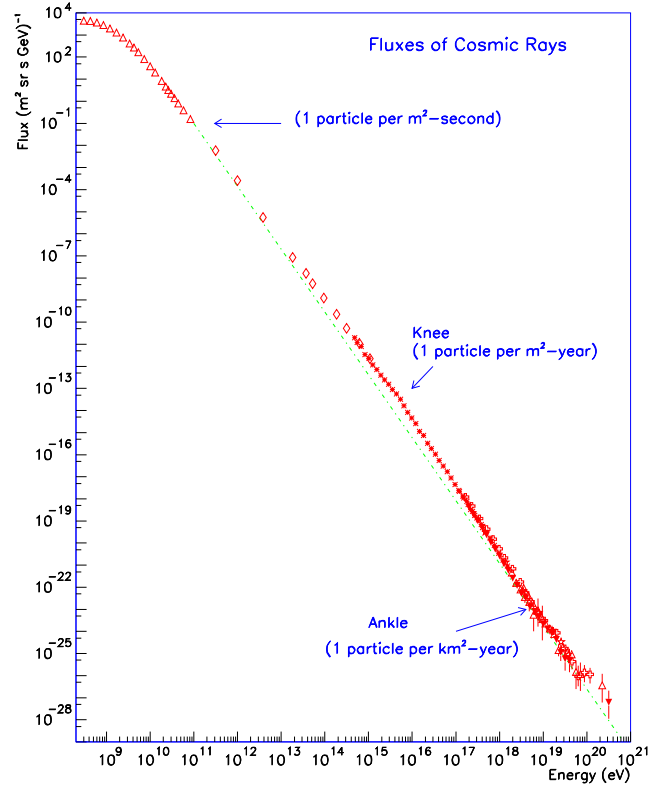,height=0.5\textheight}}}
\caption[...]{\footnotesize The cosmic ray all particle
spectrum~\cite{swordy}. Approximate integral fluxes are also shown}
\label{fig1a}
\end{figure}

After almost 90 years of research on cosmic rays (CRs), their origin
is still an open question, for which the degree of uncertainty
increases with energy: Only below 100 MeV
kinetic energy, where the solar wind shields protons coming
from outside the solar system, the sun must give rise to
the observed proton flux. The bulk of the CRs up to at least an energy
of $E=4\times 10^{15}\,$eV is believed to originate within our Galaxy.
Above that energy, which is associated with the so called ``knee'',
the flux of particles per area, time, solid angle, and energy,
which can be well approximated by broken power laws
$\propto E^{-\gamma}$, steepens from a power law index $\gamma\simeq2.7$
to one of index $\simeq3.2$. Above the so called ``ankle'' at
$E\simeq5\times10^{18}\,$eV, the spectrum flattens again to a power law
of index $\gamma\simeq2.8$. This latter feature is often interpreted as a
cross over from a steeper Galactic component to a harder component
of extragalactic origin. Fig.~\ref{fig1a} shows the measured
CR spectrum above 100 MeV, up to $3\times10^{20}\,$eV, the highest
energy measured so far for an individual CR.

The conventional scenario
assumes that all high energy charged particles are accelerated in
magnetized astrophysical shocks, whose size and typical magnetic
field strength determines the maximal achievable energy, similar
to the situation in man made particle accelerators. The most
likely astrophysical accelerators for CR up to the knee, and
possibly up to the ankle are the shocks associated with
remnants of past Galactic supernova explosions, whereas for the
presumed extragalactic component powerful objects such as
active galactic nuclei are envisaged.

\begin{figure}[htb]
\centerline{\hbox{\psfig{figure=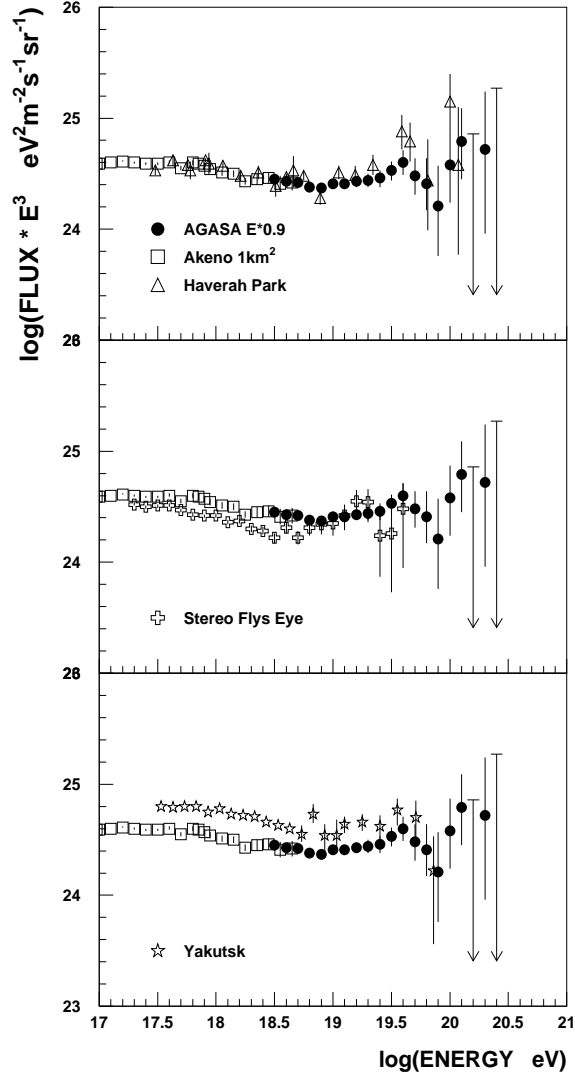,height=0.7\textheight}}}
\caption[...]{\footnotesize The cosmic ray spectrum above $10^{17}\,$eV
(from Ref~\cite{review1}).The ``ankle'' is visible at
$E\simeq5\times10^{18}\,$eV.}
\label{fig1b}
\end{figure}

The main focus of this contribution will be on ultrahigh energy
cosmic rays (UHECRs), those with energy
$\ga10^{18}\,$eV~\cite{volcano,sugar,haverah,yakutsk,fe,agasa},
see Fig.~\ref{fig1b}.
For more details on CRs in general the reader is referred
to recent monographs~\cite{bbdgp,gaisser}.
In particular, extremely high energy (EHE)\footnote{We
shall use the abbreviation EHE to specifically denote energies
$E\ga10^{20}\,$eV, while the abbreviation UHE for ``Ultra-High Energy''
will sometimes be used to denote $E\ga$ 1 EeV, where 1 EeV =
$10^{18}\,$eV. Clearly UHE includes EHE but not vice versa.}
cosmic rays pose a serious challenge for conventional theories of
CR origin based on acceleration of charged particles in powerful
astrophysical objects. The question of the origin of these EHECRs
is, therefore, currently a subject of much intense debate and
discussions as well as experimental efforts; see
Refs.~\cite{icrr90,icrr96,owl-proc}, and
Ref.~\cite{review1,reviews} for recent brief reviews, and Ref.~\cite{bs-rev}
for a detailed review. In Sect.~2 we will summarize detection
techniques and present and future experimental projects.

The current theories of origin of EHECRs can be broadly categorized
into two distinct ``scenarios'': the ``bottom-up'' acceleration
scenario, and the ``top-down'' decay scenario, with various different models
within each scenario. As the names suggest, the
two scenarios are in a sense exact opposite of each other. The
bottom-up scenario is just an extension of the conventional
shock acceleration scenario in which charged particles are
accelerated from lower energies to the requisite high energies
in certain special astrophysical environments.
On the other hand, in the top-down scenario, the energetic
particles arise simply from decay of certain sufficiently massive
particles originating from physical processes in the early
Universe, and no acceleration mechanism is needed.

The problems encountered in trying to explain EHECRs in terms of
acceleration mechanisms have been well-documented in a number of studies;
see, e.g., Refs.~\cite{hillas-araa,ssb,norman}. Even if it is
possible, in principle, to accelerate particles to EHECR energies of order
100 EeV in some astrophysical sources, it is generally extremely difficult
in most cases to get the particles come out of the dense regions in and/or 
around the sources without losing much energy. Currently, the most
favorable sources in this
regard are perhaps a class of powerful radio galaxies (see, e.g., 
Refs.~\cite{bier-rev,kirk-duffy} for recent reviews and 
references to the literature), although the values of the relevant parameters 
required for acceleration to energies $\ga$ 100 EeV are somewhat on the
extreme side~\cite{norman}. However, even
if the requirements of energetics are met, the main problem with radio
galaxies as sources of EHECRs is that most of them seem to 
lie at large cosmological distances, $\gg$ 100 Mpc, from Earth. 
This is a major problem if EHECR
particles are conventional particles such as nucleons or heavy nuclei. The
reason is that nucleons above $\simeq$ 70 EeV lose energy drastically
during their propagation from the source to Earth due to the
Greisen-Zatsepin-Kuzmin (GZK) effect~\cite{greisen,zat-kuz}, namely, 
photo-production of pions when the nucleons collide with photons of the
cosmic microwave background (CMB), the mean-free path for which is $\sim$ 
few Mpc~\cite{stecker-gzk}. This process limits the possible
distance of any source of EHE nucleons to $\la$ 100 Mpc.
If the particles were heavy nuclei, 
they would be photo-disintegrated~\cite{psb,new_heavy} in
the CMB and infrared (IR) background within similar 
distances. Thus, nucleons or heavy
nuclei originating in distant radio galaxies are unlikely to survive with
EHECR energies at Earth with any significant flux, even if they were 
accelerated to energies of order 100 EeV at source. In addition,
if cosmic magnetic fields are not close to existing
upper limits, EHECRs are not likely to be deflected strongly
by large scale cosmological and/or Galactic magnetic fields
Thus, EHECR arrival directions should point back to their
sources in the sky (see Sect.~5 for details) and EHECRs
may offer us the unique opportunity of doing
charged particle astronomy. Yet, for the observed EHECR events so far, no
powerful sources close to the arrival directions of individual events
are found within about 100 Mpc~\cite{elb-som,ssb}. Very recently,
it has been suggested by Boldt and Ghosh~\cite{boldt-ghosh} that particles 
may be accelerated to energies $\sim10^{21}\,$eV near the event horizons of
spinning supermassive black holes associated with presently {\it inactive}
quasar remnants whose numbers within the local cosmological Universe
(i.e., within a GZK distance of order 50 Mpc) may be sufficient to explain
the observed EHECR flux. This would solve the problem of absence of
suitable currently {\it active} sources associated with EHECRs. A detailed
model incorporating this suggestion, however, remains to be worked out.

There are, of course, ways to avoid the distance restriction imposed
by the GZK effect, provided
the problem of energetics is somehow solved separately and provided one
allows new physics beyond the Standard Model of particle physics; we shall
discuss those suggestions in Sect.~3.

On the other hand, in the top-down scenario, which will be discussed
in Sect.~4, the problem of energetics is
trivially solved from the beginning. Here, the EHECR particles owe their
origin to decay of some supermassive ``X'' particles of mass
$m_X\gg10^{20}\,$eV, so that their decay products, envisaged as
the EHECR particles, can have energies all the way up to $\sim m_X$. Thus,
no acceleration mechanism is needed. The sources of the massive X
particles could be topological defects such as cosmic strings or magnetic
monopoles that could be produced in the early Universe during 
symmetry-breaking phase transitions envisaged in Grand Unified
Theories (GUTs). In an inflationary early Universe, the relevant
topological defects could be formed at a phase transition at the end of
inflation. Alternatively, the X particles could be certain
supermassive metastable relic particles of lifetime comparable to
or larger than the age of the Universe, which could be
produced in the early Universe through, for example, particle production
processes associated with inflation. Absence of nearby powerful
astrophysical objects such as AGNs or radio galaxies is not a problem in
the top-down scenario because the X particles or their sources need not
necessarily be associated with any specific active astrophysical objects.
In certain models, the X particles
themselves or their sources may be clustered in galactic halos, in
which case the dominant contribution to the EHECRs observed at Earth would
come from the X particles clustered within our Galactic Halo, for which 
the GZK restriction on source distance would be of no concern. 

By focusing primarily on ``non-conventional'' scenarios
involving new particle physics beyond the electroweak scale, we
do not wish to give the wrong impression that these scenarios explain all
aspects of EHECRs. In fact, as we shall see below, essentially each of the 
specific models that have been studied so
far has its own peculiar set of problems. Indeed, the main problem of
non-astrophysical solutions of the EHECR problem in general is that
they are highly model dependent.
On the other hand, it is precisely because of this
reason that these scenarios are also attractive --- they bring
in ideas of new physics beyond the Standard Model of particle physics
(such as Grand Unification and new interactions beyond the reach of
terrestrial accelerators) as well as ideas of early Universe cosmology
(such as topological defects and/or massive particle production in
inflation) into the realms of EHECRs where these ideas have the
potential to be tested by future EHECR experiments. 

The physics and astrophysics of UHECRs are intimately linked with
the emerging field of neutrino astronomy (for reviews see
Refs.~\cite{ghs,halzen}) as well as with the already
established field of $\gamma-$ray astronomy (for reviews see, e.g.,
Ref.~\cite{gammarev}) which in turn are important
subdisciplines of particle astrophysics (for a review see, e.g.,
Ref.~\cite{mannheim}). Indeed, as we shall see, all
scenarios of UHECR origin, including the top-down models, are severely
constrained by neutrino and $\gamma-$ray observations and limits.
In turn, this linkage has important consequences for theoretical
predictions of fluxes of extragalactic neutrinos above a TeV
or so whose detection is a major goal of next-generation
neutrino telescopes (see Sect.~2): If these neutrinos are
produced as secondaries of protons accelerated in astrophysical
sources and if these protons are not absorbed in the sources,
but rather contribute to the UHECR flux observed, then
the energy content in the neutrino flux can not be higher
than the one in UHECRs, leading to the so called Waxman Bahcall
bound~\cite{wb-bound,mpr}. If one of these assumptions
does not apply, such as for acceleration sources that are opaque
to nucleons or in the TD scenarios where X particle decays
produce much fewer nucleons than $\gamma-$rays and neutrinos,
the Waxman Bahcall bound does not apply, but the neutrino
flux is still constrained by the observed diffuse $\gamma-$ray
flux in the GeV range (see Sect.~4.4).

Finally, in Sect.~5 we shall discuss how, apart from the unsolved
problem of the source mechanism, EHECR observations have the
potential to yield important information on Galactic and extragalactic
magnetic fields. 

\section{Present and Future UHE CR and Neutrino Experiments}
The CR primaries are shielded by the Earth's
atmosphere and near the ground reveal their existence only by
indirect effects such as ionization. Indeed, it was the height
dependence of this latter effect which lead to the discovery of
CRs by Hess in 1912. Direct observation of CR primaries is only
possible from space by flying detectors with balloons or
spacecraft. Naturally, such detectors are very limited in size
and because the differential CR spectrum is a steeply falling function of 
energy (see Fig.~\ref{fig1a}), direct observations run out of
statistics typically around a few $100\,$TeV.

Above $\sim100\,$TeV, the showers of secondary particles
created in the interactions of the primary CR
with the atmosphere are extensive enough to be detectable from
the ground. In the most traditional technique, charged hadronic
particles, as well as electrons and muons in these Extensive Air
Showers (EAS) are recorded on the
ground~\cite{petrera} with standard instruments 
such as water Cherenkov detectors used in the old
Volcano Ranch~\cite{volcano} and Haverah Park~\cite{haverah}
experiments, and scintillation detectors which are
used now-a-days. Currently operating ground
arrays for UHECR EAS are the Yakutsk experiment
in Russia~\cite{yakutsk} and the Akeno
Giant Air Shower Array (AGASA) near Tokyo, Japan, which is
the largest one, covering an
area of roughly $100\,{\rm km}^2$ with about 100
detectors mutually separated by about $1\,$km~\cite{agasa}.
The Sydney University Giant Air Shower Recorder (SUGAR)~\cite{sugar}
operated until 1979 and was the largest array in the Southern
hemisphere. The ground array technique allows one to
measure a lateral cross section of the shower profile.
The energy of the shower-initiating primary particle is estimated by
appropriately parametrizing it in terms of a measurable parameter;
traditionally this parameter is taken to be the particle density at 600 m
from the shower core, which
is found to be quite insensitive to the primary composition
and the interaction model used to simulate air
showers.

The detection of secondary photons from EAS represents a
complementary technique. The experimentally most important light
sources are the fluorescence of air nitrogen excited by the charged
particles in the EAS and the Cherenkov radiation from the charged
particles that travel faster than the speed of light in the
atmospheric medium. The first
source is practically isotropic whereas the second one produces
light strongly concentrated on the surface of a cone around the
propagation direction of the charged source. The fluorescence
technique can be used equally well for both charged and neutral
primaries and was first used by the Fly's Eye detector~\cite{fe}
and will be part of several future projects on UHECRs
(see below). The primary energy can be estimated from
the total fluorescence yield. Information on the primary
composition is contained in the column depth $X_{\rm max}$ 
(measured in g$\,{\rm cm}^{-2}$) at which the shower reaches maximal
particle density. The average of $X_{\rm max}$ is related to the primary
energy $E$ by
\begin{equation}
  \left\langle X_{\rm max}\right\rangle = X_0^\prime\,\ln
  \left(\frac{E}{E_0}\right)\,.\label{elongation}
\end{equation}
Here, $X_0^\prime$ is called the elongation rate and $E_0$
is a characteristic energy that depends on the primary composition.
Therefore, if $X_{\rm max}$ and $X_0^\prime$ are  
determined from the longitudinal shower profile measured
by the fluorescence detector, then $E_0$ and thus
the composition, can be extracted after determining the energy
$E$ from the total fluorescence yield. 
Comparison of CR spectra measured with the ground array
and the fluorescence technique indicate systematic errors in
energy calibration that are generally smaller than $\sim$ 40\%.
For a more detailed discussion of experimental EAS analysis with the
ground array and the fluorescence technique see, e.g.,
Refs.~\cite{easrefs}.

As an upscaled version of the old Fly's Eye Cosmic Ray experiment, the
High Resolution Fly's Eye detector is currently under construction
at Utah, USA~\cite{hires}. Taking into account a duty cycle of about
10\% (a fluorescence detector requires clear, moonless nights),
the effective aperture of this instrument will be
$\simeq350 (1000)\,{\rm km}^2\,{\rm sr}$ at $10 (100)\,$EeV,
on average about 6 times the Fly's Eye aperture, with a threshold
around $10^{17}\,$eV. Another project utilizing the fluorescence technique
is the Japanese Telescope Array~\cite{tel_array} which is currently
in the proposal stage. If approved, its effective aperture will
be about 10 times that of Fly's Eye above $10^{17}\,$eV, and
it would also be used as a Cherenkov detector for TeV $\gamma-$ray
astrophysics.
The largest project presently under construction is the  Pierre Auger
Giant Array Observatory~\cite{auger} planned for two sites, one in
Argentina and another in the USA for maximal sky coverage. Each site
will have  a $3000\,{\rm km}^2$ ground array. The southern
site will have about 1600 particle detectors (separated by 1.5 km
each) overlooked by four fluorescence
detectors. The ground arrays will have a duty cycle of nearly 100\%,
leading to an effective aperture about 30 times as large as the AGASA
array. The corresponding cosmic ray event rate above
$10^{20}\,$eV will be about 50 events per year. About 10\% 
of the events will be detected by both the ground array
and the fluorescence component and can be used for cross
calibration and detailed EAS studies. The energy threshold will
be around $10^{18}\,$eV, with full sensitivity above $10^{19}\,$eV.

Recently NASA initiated a concept study for detecting EAS
from space~\cite{owl,owl1} by observing their fluorescence light
from an Orbiting Wide-angle Light-collector (OWL). This would
provide an increase by another factor $\sim50$ in aperture
compared to the Pierre Auger Project, corresponding to an
event rate of up to a few thousand events per year above
$10^{20}\,$eV. Similar concepts such as the Extreme Universe
Space Observatory (EUSO)~\cite{euso} which is part
of the AirWatch program~\cite{airwatch} and of which a
prototype may be tested on the International Space Station
are also being discussed. It is possible that the OWL and AirWatch
efforts will merge. The energy threshold of such instruments
would be between $10^{19}$ and $10^{20}\,$eV. This technique
would be especially suitable for detection of very small
event rates such as those caused by UHE neutrinos which
would produce deeply penetrating EAS (see Sect.~4.4). For
more details on these recent experimental considerations
see Ref.~\cite{owl-proc}.

High energy neutrino astronomy is aiming towards a kilometer
scale neutrino observatory. The major technique is the optical
detection of Cherenkov light emitted by muons created in charged current
reactions of neutrinos with nucleons either in water
or in ice. The largest pilot experiments representing
these two detector media are the now defunct Deep Undersea Muon
and Neutrino Detection (DUMAND) experiment~\cite{dumand} in the
deep sea near Hawai and the Antarctic Muon And Neutrino Detector Array
(AMANDA) experiment~\cite{amanda} in the South
Pole ice. Another water based experiment is situated at
Lake Baikal~\cite{baikal}. Next generation deep sea projects
include the French Astronomy with a Neutrino Telescope and
Abyss environmental RESearch (ANTARES)~\cite{antares}
and the underwater Neutrino Experiment SouthwesT Of GReece
(NESTOR) project in the Mediterranean~\cite{nestor},
whereas ICECUBE~\cite{icecube} represents the planned kilometer scale
version of the AMANDA detector. Also under consideration are neutrino
detectors utilizing techniques to detect the radio pulse from the
electromagnetic showers created by neutrino
interactions in ice. This technique
could possibly be scaled up to an effective area of
$10^4\,{\rm km}^2$ and a prototype is represented by the
Radio Ice Cherenkov Experiment (RICE) experiment at the
South Pole~\cite{rice}. Neutrinos can also initiate
horizontal EAS which can be detected by giant ground arrays
such as the Pierre Auger Project~\cite{auger-neut}. Furthermore,
as mentioned above, deeply penetrating EAS could be detected from
space by instruments such as the proposed space based
AirWatch type detectors~\cite{owl,owl1,euso,airwatch}.
More details and references on neutrino astronomy detectors are contained
in Refs.~\cite{ghs,learned}, and some recent overviews on
neutrino astronomy can be found in Ref.~\cite{halzen}.

\section{New Primary Particles and New Interactions}

A possible way around the problem of missing counterparts
within acceleration scenarios is to propose primary
particles whose range is not limited by interactions with
the CMB. Within the Standard Model the only candidate is the neutrino,
whereas in supersymmetric extensions of the Standard Model,
new neutral hadronic bound states of light gluinos with
quarks and gluons, so-called R-hadrons that are heavier than
nucleons, and therefore have a higher GZK threshold,
have been suggested~\cite{cfk}.

In both the neutrino and new massive neutral hadron scenario
the particle propagating over extragalactic distances would have
to be produced as a secondary in
interactions of a primary proton that is accelerated in
a powerful AGN which can, in contrast to the case of
EAS induced by nucleons, nuclei, or $\gamma-$rays,
be located at high redshift. Consequently, these scenarios predict
a correlation between primary arrival directions
and high redshift sources. In fact, possible evidence
for an angular correlation 
of the five highest energy events with compact radio
quasars at redshifts between 0.3 and 2.2 was recently
reported~\cite{fb}. A new analysis with the somewhat larger data
set now available does not support significant
correlations~\cite{star}. This is currently disputed
since another group claims to have found a correlation
on the 99.9\% confidence level~\cite{virmani}.
Only a few more events could confirm or
rule out the correlation hypothesis.
Note that these scenarios would require the primary proton to
be accelerated up to $\ga10^{21}\,$eV, demanding a very powerful
astrophysical accelerator. On the other hand, a few dozen such
exceptional accelerators in the visible Universe may suffice.

\subsection{New Neutrino Interactions}

Neutrino primaries have the advantage of being well established
particles. However, within the Standard Model their interaction cross
section with nucleons, whose charged current part can
be parametrized by~\cite{gqrs}
\begin{equation}
\sigma^{SM}_{\nu N}(E)\simeq2.36\times10^{-32}(E/10^{19}
  \,{\rm eV})^{0.363}\,{\rm cm}^2\quad(10^{16}\,{\rm eV}\la
  E\la10^{21}\,{\rm eV})\,,\label{cross}
\end{equation}
falls short by about five orders of
magnitude to produce ordinary air showers.
However, it has been suggested that the neutrino-nucleon
cross section, $\sigma_{\nu N}$, can be enhanced by new
physics beyond the electroweak scale in the center of
mass (CM) frame, or above about a PeV in the nucleon rest frame.
Neutrino induced air showers may therefore rather directly
probe new physics beyond the electroweak scale.

Two major possibilities
have been discussed in the literature for which unitarity
bounds need not be violated. In the first, a broken SU(3)
gauge symmetry dual to the unbroken SU(3) color gauge group
of strong interaction is introduced as the ``generation symmetry'' such
that the three generations of leptons and quarks represent the quantum
numbers of this generation symmetry. In this scheme, neutrinos can have
close to strong interaction cross sections with quarks.
In addition, neutrinos can
interact coherently with all partons in the nucleon, resulting in
an effective cross section comparable to the geometrical
nucleon cross section.  This model lends itself to experimental
verification through shower development altitude
statistics~\cite{bhfpt}.

The second possibility consists of a large increase
in the number of degrees of freedom above the electroweak 
scale~\cite{kovesi-domokos}. A specific implementation
of this idea is given in theories with $n$ additional large
compact dimensions and a quantum gravity scale $M_{4+n}\sim\,$TeV
that has recently received much attention in the
literature~\cite{tev-qg} because it provides an alternative
solution (i.e., without supersymmetry) to the hierarchy problem
in grand unifications of gauge interactions.
The cross sections within such scenarios have not been
calculated from first principles yet. Within the field
theory approximation which should hold for squared CM
energies $s\la M_{4+n}^2$, the spin 2 character
of the graviton predicts $\sigma_g\sim s^2/M_{4+n}^6$~\cite{ns}
For $s\gg M_{4+n}^2$, several arguments based on unitarity
within field theory have been put forward. Ref.~\cite{ns}
suggested
\begin{equation}
  \sigma_{g}\simeq\frac{4\pi s}{M^4_{4+n}}\simeq
  10^{-27}\left(\frac{M_{4+n}}{{\rm TeV}}\right)^{-4}
  \left(\frac{E}{10^{20}\,{\rm eV}}\right)\,{\rm cm}^2\,,
  \label{sigma_graviton}
\end{equation}
where in the last expression we specified to a neutrino
of energy $E$ hitting a nucleon at rest. A more detailed
calculation taking into account scattering on individual
partons leads to similar orders of magnitude~\cite{jain}.
Note that a neutrino
would typically start to interact in the atmosphere
for $\sigma_{\nu N}\ga10^{-27}\,{\rm cm}^2$, i.e. in the
case of Eq.~(\ref{sigma_graviton}) for
$E\ga10^{20}\,$eV, assuming $M_{4+n}\simeq1\,$TeV.
The neutrino therefore becomes a primary candidate for the
observed EHECR events. However, since in a neutral current
interaction the neutrino transfers only about $10\%$ of its energy to
the shower, the cross section probably has to be at least a
few $10^{-26}\,{\rm cm}^2$ to be consistent with observed
showers which start within the first $50\,{\rm g}\,{\rm cm}^{-2}$
of the atmosphere~\cite{agmprs}.
A specific signature of this scenario
would be the absence of any events above the energy where
$\sigma_g$ grows beyond $\simeq10^{-27}\,{\rm cm}^2$ in
neutrino telescopes based on ice or water as detector
medium~\cite{halzen}, and a hardening of the spectrum above this energy
in atmospheric detectors such as the Pierre Auger
Project~\cite{auger} and the proposed space based AirWatch type
detectors~\cite{owl,owl1,euso,airwatch}.
Furthermore, according to Eq.~(\ref{sigma_graviton}),
the average atmospheric column depth of the first interaction
point of neutrino induced EAS in this scenario is predicted
to depend linearly on energy. This should be easy to distinguish
from the logarithmic scaling, Eq.~(\ref{elongation}), expected
for nucleons, nuclei, and $\gamma-$rays. To test
such scalings one can, for example, take advantage of the
fact that the atmosphere provides a detector medium whose
column depth increases from $\sim1000\,{\rm g}/{\rm cm}^2$
towards the zenith to $\sim36000\,{\rm g}/{\rm cm}^2$
towards horizontal arrival directions. This probes
cross sections in the range $\sim10^{-29}-10^{-27}\,{\rm cm}^2$.
Due to the increased Water/ice detectors would probe
cross sections in the range
$\sim10^{-31}-10^{-29}\,{\rm cm}^2$~\cite{tol}.

Within string theory, individual amplitudes are expected
to be suppressed exponentially above the string scale $M_s$
which for simplicity we assume here to be comparable to $M_{4+n}$.
This can be interpreted as a result of the finite spatial
extension of the string states.
In this case, the neutrino nucleon cross section would
be dominated by interactions with the partons carrying
a momentum fraction $x\sim M_s^2/s$, leading to~\cite{kp}
\begin{eqnarray}
  \sigma_{\nu N}&\simeq&\frac{4\pi}{M_s^2}\ln(s/M_s^2)
  (s/M_s^2)^{0.363}\nonumber\\
  &\simeq&6\times10^{-29}\left(\frac{M_s}{{\rm TeV}}\right)^{-4.726}
  \left(\frac{E}{10^{20}\,{\rm eV}}\right)^{0.363}\label{sigma_nuN}\\
  &&\times\left[1+0.08\ln\left(\frac{E}{10^{20}\,{\rm eV}}\right)
  -0.16\ln\left(\frac{M_s}{{\rm TeV}}\right)\right]^2\,{\rm cm}^2
  \nonumber
\end{eqnarray}
This is probably too small to make neutrinos primary candidates
for the highest energy showers observed, given the
fact that complementary constraints from accelerator
experiments result in $M_s\ga 1,\,$TeV~\cite{cpp}.
On the other hand, in the total cross section amplitude
suppression may be compensated by an exponential growth of
the level density~\cite{kovesi-domokos}. It is currently
unclear and it may be model dependent which effect dominates.
Thus, an experimental detection of the signatures discussed
in this section could lead to constraints on some
string-inspired models of extra dimensions.

We note in passing that extra dimensions can have other
astrophysical ramifications such as energy loss in stellar
environments due to emission of real gravitons into the
bulk. The strongest resulting lower limits on $M_{4+n}$
come from the consideration of cooling of the cores of
hot supernovae and read $M_6\ga50\,$TeV, $M_7\ga4\,$TeV,
$M_8\ga1\,$TeV, and $M_{11}\ga0.05\,$TeV for $n=2,3,4,7$,
respectively~\cite{sn87a}. In addition, implications
of extra dimensions for early Universe physics and inflation
are increasingly studied in the literature, but much
work is left to be done on the intersection of these
research domains.

Independent of theoretical arguments, the UHECR data
can be used to put constraints on cross sections
satisfying $\sigma_{\nu N}(E\ga10^{19}\,{\rm eV})
\la10^{-27}\,{\rm cm}^2$. Particles with such cross
sections would give rise to horizontal air showers.
The Fly's Eye experiment established an upper limit
on horizontal air showers~\cite{baltrusaitis}. The non-observation of
the neutrino flux expected from pions produced by
UHECR interacting with the CMB the results in the
limit~\cite{tol}
\begin{eqnarray}
  \sigma_{\nu N}(10^{17}\,{\rm eV})&\la&1\times10^{-29}
  /{\bar y}^{1/2}\,{\rm cm}^2\nonumber\\
  \sigma_{\nu N}(10^{18}\,{\rm eV})&\la&8\times10^{-30}
  /{\bar y}^{1/2}\,{\rm cm}^2\nonumber\\
  \sigma_{\nu N}(10^{19}\,{\rm eV})&\la&5\times10^{-29}
  /{\bar y}^{1/2}\,{\rm cm}^2\,,\label{crosslim1}
\end{eqnarray}
where ${\bar y}$ is the average energy fraction of the neutrino
deposited into the shower (${\bar y}=1$ for charged current
reactions and ${\bar y}\simeq0.1$ for neutral current reactions).
Expected neutrino fluxes are shown in Fig.~\ref{fig5}.
The projected sensitivity of future experiments such as
the Pierre Auger Observatories and the AirWatch type satellite
projects indicate that the cross section limits Eq.~(\ref{crosslim1})
could be improved by up to four orders of magnitude,
corresponding to one order of magnitude in $M_s$ or
$M_{4+n}$.

\subsection{Supersymmetric Particles}
Light gluinos binding to quarks, anti-quarks and/or gluons
can occur in supersymmetric theories 
involving gauge-mediated supersymmetry (SUSY) breaking~\cite{raby} where
the resulting gluino mass arises dominantly from radiative corrections
and can vary between $\sim1\,$GeV and $\sim100\,$GeV.
In these scenarios, the gluino can be the lightest supersymmetric
particle (LSP). There are also arguments against a light
quasi-stable gluino~\cite{vo}, mainly based on constraints on
the abundance of anomalous heavy isotopes of hydrogen and oxygen
which could be formed as bound states of these nuclei and the
gluino. Furthermore, accelerator constraints have become
quite stringent~\cite{acc_limits} and seem to be inconsistent with the
original scenario from Ref.~\cite{cfk}. However, the scenario with
a ``tunable'' gluino mass~\cite{raby} still seems possible
and suggests either the gluino--gluon bound state $g\tilde g$,
called glueballino $R_0$, or the isotriplet $\tilde g-(u\bar u-
d\bar d)_8$, called $\tilde\rho$, as the lightest quasi-stable
R-hadron. For a summary of scenarios with light gluinos
consistent with accelerator constraints see Ref.~\cite{clavelli}.
The case of a light quasi-stable gluino does
not seem to be settled.

\begin{figure}[htb]
\centerline{\hbox{\psfig{figure=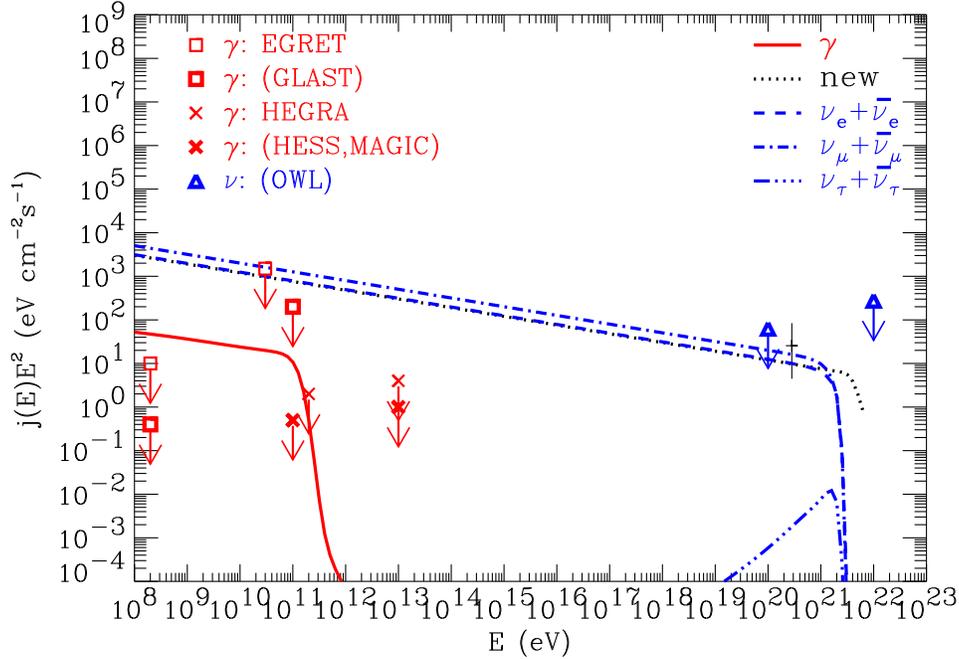,height=0.4\textheight}}}
\caption[...]{\footnotesize Schematic predictions for the fluxes of the
putative new neutral heavy particle (dotted
line), electron, muon, and $\tau-$neutrinos (dashed and dash-dotted
lines, as indicated), and $\gamma-$rays (solid line) for a
source at redshift $z=1$. Assumed were
a proton spectrum $\propto E^{-2.2}$ extending at least
up to $10^{22}\,$eV at the source, a branching ratio
for production of the heavy neutral in nucleon interactions
of 0.01, and a beaming factor of 10 for neutrinos and the
heavy neutrals. The 1 sigma error bar at $3\times10^{20}\,$eV
represents the point flux corresponding to the highest
energy Fly's Eye event. The predicted fluxes were normalized
such that this highest energy event is explained as a
new heavy particle. The points with arrows on the right part
represent projected approximate neutrino point source sensitivities
for the space based AirWatch type concepts using the OWL/AirWatch
acceptance estimated in Ref.~\cite{owl,owl1}
for non-detection over a five year period.
The points with arrows in the lower left part
represent approximate $\gamma-$ray point source sensitivities of
existing detectors such as EGRET and HEGRA,
and of planned instruments such as the satellite detector
GLAST, the Cherenkov telescope array HESS
and the single dish instrument MAGIC, for 50 hours
and 1 month observation time for the ground based and satellite
detectors, respectively.}
\label{fig2}
\end{figure}

An astrophysical constraint on new neutral massive and strongly 
interacting EAS primaries results from the fact that the
nucleon interactions producing these particles in the source
also produce neutrinos and especially $\gamma-$rays. The
resulting fluxes from powerful discrete acceleration sources may be easily
detectable in the GeV range by space-borne $\gamma-$ray instruments
such as EGRET and GLAST, and in the TeV range by ground based
$\gamma-$ray detectors such as HEGRA and WHIPPLE and the planned
VERITAS, HESS, and MAGIC projects (for reviews discussing these
instruments see Ref.~\cite{gammarev}). At least
the latter three ground based instruments should have energy thresholds low
enough to detect $\gamma-$rays from the postulated sources
at redshift $z\sim1$.
Such observations in turn imply constraints on the required
branching ratio of proton interactions into the R-hadron which,
very roughly, should be larger than $\sim0.01$. These
constraints, however, will have to be investigated in more
detail for specific sources. One could also search
for heavy neutral baryons in the data from Cherenkov
instruments in the TeV range in this context. To demonstrate
these  points, a schematic
example of fluxes predicted for the new heavy particle and
for $\gamma-$rays and neutrinos are shown in Fig.~\ref{fig2}.

A further constraint on new EAS primary
particles in general comes from the character of the air
showers created by them: The observed EHECR air showers
are consistent with nucleon primaries and limits the
possible primary rest mass to less than
$\simeq50\,$GeV~\cite{afk}. With the statistics expected
from upcoming experiments such as the Pierre Auger Project,
this upper limit is likely to be lowered down to
$\simeq10\,$GeV.

It is interesting to note in this context that in case of
a confirmation of the existence of new neutral particles
in UHECRs, a combination of accelerator, air shower, and
astrophysics data would be highly restrictive in terms of
the underlying physics: In the above scenario, for example,
the gluino would have to be in a narrow mass range, 1--10
GeV, and the newest accelerator constraints on the Higgs
mass, $m_h\ga90\,$GeV, would require the presence of
a D term of an anomalous $U(1)_X$ gauge symmetry, in
addition to a gauge-mediated contribution to SUSY breaking
at the messenger scale~\cite{raby}.

\subsection{Anomalous Kinematics, Quantum Gravity Effects,
Lorentz Symmetry Violations}
The existence of UHECR beyond the GZK cutoff has prompted several
suggestions of possible new physics beyond the Standard Model. We have
already discussed some of these suggestions in Sect.~3.1
and~3.2 in the context of new primary particles and
new interactions. Further,   
in Sect.~4 we will discuss suggestions regarding possible
new sources of EHECR that also involve postulating new physics beyond the
Standard Model. In the present section, to end our discussions on the
propagation and new interactions
of UHE radiation, we briefly discuss some examples
of possible small violations or modifications of certain fundamental
tenets of physics (and constraints on the magnitude of those
violations/modifications) that have also been discussed in the literature
in the context of propagation of UHECR. 

For example, as an interesting consequence of the very existence of UHECR,
constraints on possible violations of Lorentz invariance (VLI)  
have been pointed out~\cite{colgla}. These constraints rival
precision measurements in
the laboratory: If events observed around $10^{20}\,$eV are indeed
protons, then the difference between the maximum attainable proton
velocity and the speed of light has to be less than about 
$1\times10^{-23}$, otherwise the proton would lose its energy by
Cherenkov radiation within a few hundred centimeters. Possible
tests of other modes of VLI with UHECR
have been discussed in Ref.~\cite{g-m}, and in Ref.~\cite{cowsik-sreek} in
the context of horizontal air-showers generated by cosmic rays in general.  
Gonzalez-Mestres~\cite{g-m}, Coleman and Glashow~\cite{colgla2}, and
earlier, Sato and Tati~\cite{sato-tati} and Kirzhnits and
Chechin~\cite{kirzh-che} have also suggested that due to modified
kinematical constraints the GZK cutoff could even be evaded by allowing
a tiny VLI too small to have been
detected otherwise. Similar consequences apply to other
energy loss processes such as pair production by photons above
a TeV with the low energy photon background~\cite{kifune}.
It seems to be possible to accomodate such effects within
theories involving generalized Lorentz transformations~\cite{bogo-goenner}
or deformed relativistic kinematics~\cite{g-m1}.
Furthermore, it has been pointed out~\cite{halprin-kim} that violations of
the principle of equivalence (VPE), while not dynamically equivalent, also
produce the same kinematical effects as VLI for particle processes 
in a constant gravitational potential, and so the constraints on VLI  
from UHECR physics can be translated into constraints
on VPE such that the difference between the couplings of protons and
photons to gravity must be less than about $1\times10^{-19}$. Again, this 
constraint is more stringent by several orders of magnitude than
the currently available laboratory constraint from E\"{o}tv\"{o}s
experiments. 

As a specific example of VLI, we consider an energy dependent photon group
velocity $\partial E/\partial k=c[1-\chi E/E_0+{\cal O}(E^2/E_0^2)]$
where $c$ is the speed of light in the low energy limit, $\chi=\pm1$,
and $E_0$ denotes the energy scale where this modification
becomes of order unity. This corresponds to a dispersion relation
\begin{equation}
  c^2k^2\simeq E^2+\chi\frac{E^3}{E_0}\,,\label{mod-dispersion}
\end{equation}
which, for example, can occur in quantum gravity and string
theory~\cite{emn}. The kinematics of electron-positron pair production
in a head-on collision of a high energy photon of energy $E$ with a
low energy background photon of energy $\varepsilon$ then leads to the
constraint
\begin{equation}
  \varepsilon\simeq\frac{E}{4}\left(\frac{m_e^2}{E_1E_2}+
  \theta_1\theta_2\right)+\chi\frac{E^2}{4E_0}\,,\label{mod-kinematics}
\end{equation}
where $E_i$ and $\theta_i\sim{\cal O}(m/E_i)$ are respectively the energy
and outgoing momentum angle (with respect to the original photon
momentum) of the electron and positron ($i=1,2$). 
For the case considered by Coleman and Glashow~\cite{colgla} in which the
maximum attainable speed $c_i$ of the matter particle is different
from the photon speed $c$, the kinematics can be obtained by
substituting $c_i^2-c^2$ for $\chi E/E_0$ in Eq.~(\ref{mod-kinematics}).

Let us define a critical energy $E_c=(m_e^2E_0)^{1/3}
\simeq15(E_0/m_{\rm Pl})^{1/3}\,$TeV in the case of the energy dependent
photon group velocity, and  
$E_c=m_e/|c_i^2-c^2|^{1/2}$ in the case considered by Coleman and
Glashow.
If $\chi<0$, or $c_i<c$, then $\varepsilon$ becomes negative for
$E\ga E_c$. This signals that the photon can spontaneously
decay into an electron-positron pair and propagation of photons across
extragalactic distances 
will in general be inhibited. The observation of extragalactic
photons up to $\simeq20\,$TeV~\cite{mrk421,mrk501} therefore puts the limits
$E_0\ga M_{\rm Pl}$ or $c_i^2-c^2\ga-2\times10^{-17}$.
In contrast, if $\chi>0$, or $c_i>c$,
$\varepsilon$ will grow with energy for $E\ga E_c$ until
there is no significant number of target photon density available
and the Universe becomes transparent to UHE photons.
A clear test of this possibility would be the observation of
$\ga100\,$TeV photons from distances
$\ga100\,$Mps~\cite{kluzniak}. An unambiguous detection of
the thresholds for pion production by nucleons and for
pair production by photons in the CMB and other low energy
photon backgrounds in the future would allow to establish
more stringent lower limits on $E_0$~\cite{abgg}.

In addition, the dispersion relation Eq.~(\ref{mod-dispersion})
implies that a photon signal at energy $E$ will be spread out
by $\Delta t\simeq (d/c)(E/E_0)\simeq1(d/100\,{\rm Mpc})(E/\,{\rm TeV})
(E_0/M_{\rm Pl})^{-1}\,$s. The observation of $\gamma-$rays
at energies $E\ga2\,$TeV within $\simeq300\,$s from the AGN Markarian
421 therefore puts a limit (independent of $\chi$) of
$E_0\ga4\times10^{16}\,$GeV, whereas the possible observation
of $\gamma-$rays at $E\ga200\,$TeV within $\simeq200\,$s from a
GRB by HEGRA might be sensitive to $E_0\simeq M_{\rm Pl}$~\cite{acemns}.
For a recent detailed discussion of these limits see Ref.~\cite{efmmn}.

A related proposal originally due to Kosteleck\'{y}
in the context of CR suggests the electron neutrino to be a
tachyon~\cite{kostelecky}.
This would allow the proton in a nucleus of mass $m(A,Z)$ for mass
number $A$ and charge $Z$ to decay via $p\to n+e^++\nu_e$ above
the energy threshold $E_{th}=m(A,Z)[m(A,Z\pm1)+m_e-m(A,Z)]/|m_{\nu_e}|$
which, for a free proton, is $E_{th}\simeq1.7\times10^{15}/
(|m_{\nu_e}|/{\rm eV})\,{\rm eV}$. Ehrlich~\cite{ehrlich} claims
that by choosing $m^2_{\nu_e}\simeq-(0.5\,{\rm eV})^2$ it is
possible to explain the knee and several other features
of the observed CR spectrum, including the high energy end,
if certain assumptions about the source distribution are made.
The experimental best fit values of $m^2_{\nu_e}$ from tritium beta
decay experiments are indeed negative~\cite{p-data}, although this is most
likely due to unresolved experimental issues. In addition, the values of 
$|m^2_{\nu_e}|$ from tritium beta decay experiments are typically larger
than the value required to fit the knee of the CR spectrum. This scenario
also predicts a neutron line around the knee energy~\cite{ehrlich1}.

\section{Top-Down Scenarios}

\subsection{The Main Idea}

As mentioned in the introduction, all top-down scenarios
involve the decay of X particles of mass close to the GUT scale
which can basically be produced in two ways: If they are very
short lived, as usually expected in many GUTs, they have to be
produced continuously. The only way this can be achieved is
by emission from topological defects left over from cosmological
phase transitions that may have occurred in the early Universe at
temperatures close to the GUT scale, possibly during reheating
after inflation. Topological defects
necessarily occur between regions that are causally disconnected, such
that the orientation of the order parameter
associated with the phase transition can not be communicated
between these regions and consequently will adopt different
values. Examples are cosmic strings (similar to vortices in superfluid
helium), magnetic monopoles, and domain walls (similar to Bloch
walls separating regions of different magnetization in a ferromagnet).
The defect density is consequently given by the particle horizon
in the early Universe and their formation can even be studied
in solid state experiments where the expansion rate of the Universe
corresponds to the quenching speed with which the phase
transition is induced~\cite{vachaspati}. The defects are
topologically stable, but
in the cosmological case time dependent motion
leads to the emission of particles with a mass comparable to the
temperature at which the phase transition took place. The
associated phase transition can also occur during reheating
after inflation.

Alternatively, instead of being released from topological
defects, X particles
may have been produced directly in the early Universe and,
due to some unknown symmetries, have a very
long lifetime comparable to the age of the Universe.
In contrast to Weakly-Interacting Massive Particles (WIMPS)
below a few hundred TeV which are the usual dark matter
candidates motivated by, for example, supersymmetry and can
be produced by thermal freeze out, such superheavy X particles
have to be produced non-thermally.
Several such mechanisms operating in the
post-inflationary epoch in the early Universe have been studied. They
include gravitational production through the effect of the expansion of
the background metric on the vacuum quantum fluctuations of the
X particle field, or creation during reheating at
the end of inflation if the X particle field couples to the inflaton
field. The latter case can be divided into
three subcases, namely ``incoherent'' production with an
abundance proportional to the X particle annihilation cross section,
non-adiabatic production in broad parametric resonances with
the oscillating inflaton field during preheating (analogous
to energy transfer in a system of coupled pendula), and creation
in bubble wall collisions if inflation is completed by a first
order phase transition. In all these cases, such particles,
also called ``WIMPZILLAs'', would contribute to the dark matter
and their decays could still contribute to UHE CR fluxes today,
with an anisotropy pattern that reflects the dark matter
distribution in the halo of our Galaxy.

It is interesting to note that one of the prime motivations
of the inflationary paradigm was to dilute excessive production
of ``dangerous relics'' such as topological defects and
superheavy stable particles. However, such objects can be
produced right after inflation during reheating
in cosmologically interesting abundances, and with a mass scale
roughly given by the inflationary scale which in turn
is fixed by the CMB anisotropies to
$\sim10^{13}\,$GeV~\cite{kuz-tak}. The reader will realize that
this mass scale is somewhat above the highest energies
observed in CRs, which implies that the decay products of
these primordial relics could well have something to do with
EHECRs which in turn can probe such scenarios!

For dimensional reasons the spatially averaged X particle
injection rate can only
depend on the mass scale $m_X$ and on cosmic time $t$ in the
combination
\begin{equation}
  \dot n_X(t)=\kappa m_X^p t^{-4+p}\,,\label{dotnx}
\end{equation}
where $\kappa$ and $p$ are dimensionless constants whose
value depend on the specific top-down scenario~\cite{BHS},
For example, the case $p=1$ is representative of scenarios
involving release of X particles from topological defects,
such as ordinary cosmic
strings~\cite{BR}, necklaces~\cite{BV} and magnetic 
monopoles~\cite{BS}. This can be easily seen as follows:
The energy density $\rho_s$ in a network of defects has to scale
roughly as the critical density, $\rho_s\propto\rho_{\rm crit}\propto
t^{-2}$, where $t$ is cosmic time, otherwise the defects
would either start to overclose the Universe, or end up
having a negligible contribution to the total energy
density. In order to maintain this scaling, the defect
network has to release energy with a rate given by
$\dot\rho_s=-a\rho_s/t\propto t^{-3}$, where $a=1$ in
the radiation dominated aera, and $a=2/3$ during matter
domination. If most of this energy goes into emission
of X particles, then typically $\kappa\sim{\cal O}(1)$.
In the numerical simulations presented below, it was
assumed that the X particles are nonrelativistic at decay.

The X particles could be gauge bosons, Higgs bosons, superheavy fermions,
etc.~depending on the specific GUT. They would have
a mass $m_X$ comparable to the symmetry breaking scale and would
decay into leptons and/or quarks of roughly
comparable energy. The quarks interact strongly and 
hadronize into nucleons ($N$s) and pions, the latter
decaying in turn into $\gamma$-rays, electrons, and neutrinos. 
Given the X particle production rate, $dn_X/dt$, the effective
injection spectrum of particle species $a$ ($a=\gamma,N,e^\pm,\nu$) 
via the hadronic channel can be
written as $(dn_X/dt)(2/m_X)(dN_a/dx)$,
where $x \equiv 2E/m_X$, and $dN_a/dx$ is the relevant
fragmentation function (FF).

We adopt the Local Parton Hadron Duality (LPHD) approximation~\cite{detal}
according to which the total
hadronic FF, $dN_h/dx$, is taken to be proportional to the spectrum
of the partons (quarks/gluons) in the parton cascade (which is initiated
by the quark through perturbative QCD processes) after evolving the parton
cascade to a stage where the typical transverse momentum transfer in the
QCD cascading processes has come down to $\sim R^{-1}\sim$ few hundred 
MeV, where $R$ is a typical hadron size. The parton spectrum is obtained
from solutions of the standard QCD evolution equations in modified leading
logarithmic approximation (MLLA) which provides good fits to accelerator
data at LEP energies~\cite{detal}. We will specifically use a recently
suggested generalization of the MLLA spectrum that includes the effects 
of supersymmetry~\cite{BK}. Within the LPHD hypothesis, the pions
and nucleons after hadronization have essentially the same spectrum. 
The LPHD does not, however, fix the relative abundance of pions and
nucleons after hadronization. Motivated by accelerator data, we assume
the nucleon content $f_N$ of the hadrons to be in the range
3 to 10\%, and the rest pions distributed
equally among the three charge states. According to recent Monte Carlo
simulations~\cite{BirSar}, the nucleon-to-pion ratio may
be significantly higher in certain ranges of $x$ values at the 
extremely high energies of interest here. Unfortunately, however, 
due to the very nature of these Monte Carlo calculations, it is difficult
to understand the precise physical reason for the unexpectedly high baryon
yield relative to mesons. While more of these Monte Carlo calculations of
the relevant FFs in the future will hopefully clarify the situation, we
will use here the range of $f_N\sim$ 3 to 10\% mentioned above,
which also seems to be supported by other recent Monte Carlo
simulations~\cite{berek}.
The standard pion decay spectra then give the injection spectra
of $\gamma$-rays, electrons, and neutrinos. For more details
concerning uncertainties in the X particle decay spectra
see Ref.~\cite{SLBY}.

\subsection{Numerical Simulations}
The $\gamma$-rays and electrons produced by  X particle decay
initiate  electromagnetic
(EM) cascades on low energy radiation fields such as the
CMB. The high energy photons undergo electron-positron pair
production (PP; $\gamma \gamma_b \rightarrow e^- e^+$), and 
at energies below $\sim 10^{14}$ eV they interact mainly with 
the universal infrared and optical (IR/O) backgrounds, while above 
$\sim 100$ EeV  they interact mainly with the universal radio background (URB).
In the Klein-Nishina regime, where the CM energy is
large compared to the electron mass, one of the outgoing particles usually
carries most of the initial energy. This ``leading''
electron (positron) in turn can transfer almost all of its energy to
a background photon via inverse
Compton scattering (ICS; $e \gamma_b \rightarrow e^\prime\gamma$).
EM cascades are driven by this cycle of PP and ICS.
The energy degradation of the ``leading'' particle in this cycle
is slow, whereas the total number of particles grows
exponentially with time. This makes a standard Monte Carlo
treatment difficult. Implicit numerical schemes have therefore been
used to solve the relevant kinetic 
equations. A detailed account of the transport equation approach
used in the calculations whose results are presented in this
contribution can be found in Ref.~\cite{Lee}. All
EM interactions that influence the $\gamma$-ray spectrum in the energy range
$10^8\,{\rm eV} < E < 10^{25}\,$eV, namely PP, ICS, triplet pair
production (TPP; $e \gamma_b
\rightarrow e e^- e^+$), and double pair production (DPP, $\gamma \gamma_b
\rightarrow e^-e^+e^-e^+$), as well as synchrotron losses
of electrons in the large scale extragalactic magnetic field
(EGMF), are included.

Similarly to photons, UHE neutrinos give rise to neutrino
cascades in the primordial neutrino background via exchange
of W and Z bosons~\cite{Zburst,ydjs}. Besides the secondary
neutrinos which drive the neutrino cascade, the W and Z decay products
include charged leptons and quarks which in turn feed into the
EM and hadronic channels. Neutrino interactions become
especially significant if the relic neutrinos have masses $m_\nu$
in the eV range and thus constitute hot dark matter, because
the Z boson resonance then occurs at an UHE neutrino energy
$E_{\rm res}=4\times10^{21}({\rm eV}/m_\nu)$ eV. In fact, this has been
proposed as a significant source of EHECRs~\cite{weiler2,YSL}.
Motivated by recent experimental evidence for neutrino mass
we assumed a mass of 1 eV for all three neutrino flavors (for
simplicity) and implemented the relevant W boson interactions in the
t-channel and the Z boson exchange via t- and s-channel. Hot dark matter
is also expected to cluster, potentially increasing secondary
$\gamma$-ray and nucleon production~\cite{weiler2,YSL}. This influences
mostly scenarios where X decays into neutrinos only. We
parametrize massive neutrino clustering by a length scale $l_\nu$
and an overdensity $f_\nu$ over the average density $\bar{n_\nu}$.
The Fermi distribution with a velocity dispersion $v$ yields
$f_\nu\la v^3 m_\nu^3/(2\pi)^{3/2}/\bar{n_\nu}\simeq
330\,(v/500\,{\rm km}\,{\rm sec}^{-1})^3\,
(m_\nu/{\rm eV})^3$~\cite{peebles}. Therefore, values of
$l_\nu\simeq$ few Mpc and $f_\nu\simeq20$ are conceivable
on the local Supercluster scale~\cite{YSL}.

The relevant nucleon interactions implemented are
pair production by protons ($p\gamma_b\rightarrow p e^- e^+$),
photoproduction of single or multiple pions ($N\gamma_b \rightarrow N
\;n\pi$, $n\geq1$), and neutron decay.
In TD scenarios, the particle injection spectrum is generally dominated
by the ``primary'' $\gamma$-rays and neutrinos over nucleons. These
primary $\gamma$-rays and neutrinos are produced by the decay of
the primary pions resulting from the hadronization of quarks that come
from the decay of the X particles. The contribution of secondary
$\gamma$-rays, electrons, and neutrinos from decaying pions that are
subsequently produced by the interactions of nucleons with
the CMB, is in general negligible compared to that of the primary
particles; we nevertheless include the contribution of the
secondary particles in our code.

In principle, new interactions such as the ones involving
a TeV quantum gravity scale can not only modify the interactions
of primary particles in the detector, as discussed in Sect.~3.1,
but also their propagation. However, $\gamma-$rays and
nuclei interact mostly with the CMB and IR for which the
CM energy is at most
$\simeq30(E/10^{14}\,{\rm GeV})^{1/2}\,$GeV. At such energies
the new interactions are much weaker than the dominating
electromagnetic and strong interactions. It has been
suggested recently~\cite{neutrino_int} that new interactions
may notably influence UHE neutrino propagation. However,
for neutrinos of mass $m_\nu$, the CM energy is
$\simeq100(E/10^{14}\,{\rm GeV})^{1/2}(m_\nu/0.1\,{\rm eV})^{1/2}\,$GeV,
and therefore, UHE neutrino propagation would only be
significantly modified for neutrino masses significantly
larger than $0.1\,$eV. We will therefore ignore this possibility
here.

We assume a flat Universe with no cosmological constant,
and a Hubble constant of $h=0.65$ in units of
$100\;{\rm km}\;{\rm sec}^{-1}{\rm Mpc}^{-1}$ throughout.
The numerical calculations follow
{\it all} produced particles in the EM, hadronic,
and neutrino channel, whereas the often-used continuous energy loss (CEL)
approximation (e.g., \cite{ABS}) follows
only the leading cascade particles. The CEL approximation
can significantly underestimate the cascade flux at lower energies.

The two major uncertainties in the particle transport are the
intensity and spectrum of the URB for which there exists only
an estimate above a few MHz frequency~\cite{Clark}, and the average value
of the EGMF. To bracket these uncertainties, simulations
have been performed for the observational URB estimate from
Ref.~\cite{Clark} that
has a low-frequency cutoff at 2 MHz (``minimal''), and the medium
and maximal theoretical estimates
from Ref.~\cite{PB}, as well as for EGMFs between zero
and $10^{-9}$ G, the latter motivated by limits from
Faraday rotation measurements, see Sect.~5.2 below. A strong URB tends
to suppress the UHE $\gamma$-ray flux by direct absorption
whereas a strong EGMF blocks EM cascading (which otherwise develops
efficiently especially in a low URB) by synchrotron cooling
of the electrons. For the IR/O background we used the most
recent data~\cite{irb}.

\subsection{Results: $\gamma-$ray and Nucleon Fluxes}

\begin{figure}[htb]
\centerline{\hbox{\psfig{figure=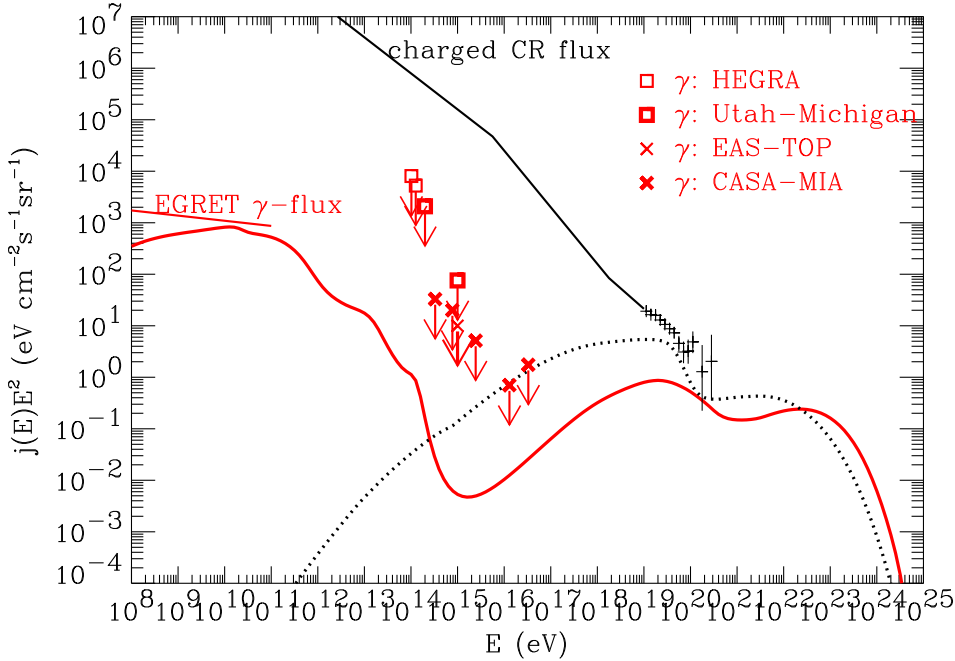,height=0.4\textheight}}}
\caption[...]{\footnotesize Predictions for the differential fluxes of
$\gamma-$rays (solid line) and protons and neutrons (dotted
line) in a TD model characterized by $p=1$, $m_X = 10^{16}\,$GeV,
and the decay mode $X\to q+q$, assuming the supersymmetric modification of
the fragmentation function~\cite{BK}, with a fraction of about
10\% nucleons. The defects have been assumed to be homogeneously
distributed. The calculation
used the code described in Ref.~\cite{SLBY} and assumed the strongest
URB version from Ref.~\cite{PB} and an EGMF $\ll10^{-11}\,$G.
1 sigma error bars are the combined data from the Haverah Park~\cite{haverah},
the Fly's Eye~\cite{fe}, and the AGASA~\cite{agasa} experiments
above $10^{19}\,$eV. Also shown are piecewise power law fits to the observed
charged CR flux (thick solid line) and the EGRET measurement
of the diffuse $\gamma-$ray flux between 30 MeV and 100 GeV~\cite{egret}
(solid line on left margin). Points with arrows
represent upper limits on the
$\gamma-$ray flux from the HEGRA, the
Utah-Michigan, the EAS-TOP, and
the CASA-MIA experiments, as indicated.
\label{fig3}}
\end{figure}

\begin{figure}[htb]
\centerline{\hbox{\psfig{figure=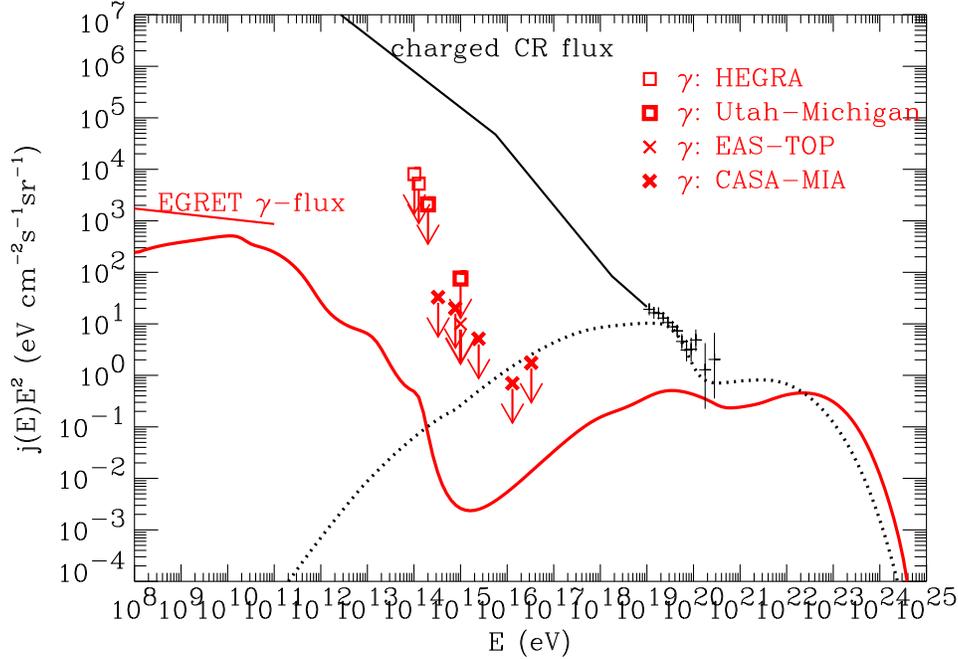,height=0.4\textheight}}}
\caption[...]{\footnotesize Same as Fig.~\ref{fig3}, but for an EGMF of
$10^{-9}\,$G.
\label{fig4}}
\end{figure}

Fig.~\ref{fig3} shows results from Ref.~\cite{SLBY} for the
time averaged $\gamma-$ray and nucleon
fluxes in a typical TD scenario, assuming no
EGMF, along with current observational constraints on
the $\gamma-$ray flux. The spectrum was optimally normalized 
to allow for an explanation of the observed EHECR
events, assuming their consistency with a nucleon or
$\gamma-$ray primary. The flux below $\la2\times10^{19}\,$eV
is presumably due to conventional
acceleration in astrophysical sources and was not fit. Similar spectral
shapes have been obtained in Ref.~\cite{ps1}, where the normalization
was chosen to match the observed differential flux at
$3\times10^{20}\,$eV. This normalization, however, 
leads to an overproduction of the
integral flux at higher energies, whereas above $10^{20}\,$eV,
the fits shown in Figs.~\ref{fig3} and~\ref{fig4} have
likelihood significances above 50\% (see Ref.~\cite{slsb} for
details) and are consistent with the integral flux above
$3\times10^{20}\,$eV estimated in Refs.~\cite{fe,agasa}.
The PP process on the CMB depletes the photon flux above 100 TeV, and the
same process on the IR/O background causes depletion of the photon flux in
the range 100 GeV--100 TeV, recycling the absorbed energies to
energies below 100 GeV through EM cascading (see Fig.~\ref{fig3}).
The predicted background is {\it not} very sensitive to
the specific IR/O background model, however~\cite{ahacoppi}.
The scenario in Fig.~\ref{fig3} obviously
obeys all current constraints within the normalization
ambiguities and is therefore quite viable. Note
that the diffuse $\gamma-$ray background measured by
EGRET~\cite{egret} up to 10 GeV puts a strong constraint on these
scenarios, especially if there is already a significant
contribution to this background from conventional
sources such as unresolved $\gamma-$ray blazars~\cite{muk-chiang}.
However, the $\gamma-$ray background constraint can be circumvented by
assuming that TDs or the decaying long lived X particles
do not have a uniform density throughout the
Universe but cluster within galaxies~\cite{bkv}. As can
also be seen, at
energies above 100 GeV, TD models are not significantly
constrained by observed $\gamma-$ray fluxes yet (see
Ref.~\cite{bs-rev} for more details on these measurements).

Fig.~\ref{fig4} shows results for the same TD scenario as in
Fig.~\ref{fig3}, but for a high EGMF $\sim 10^{-9}\,$G,
somewhat below the current upper limit, see Eq.~(\ref{newF}) below.
In this case, rapid synchrotron cooling of the initial cascade pairs quickly
transfers energy out of the UHE range. The UHE $\gamma-$ray flux then depends
mainly on the absorption length due to pair production and is typically
much lower~\cite{ABS,los}. (Note, though, that for $m_X\ga
10^{25}$ eV, the synchrotron radiation from these pairs
can be above $10^{20}\,$eV, and the UHE flux
is then not as low as one might expect.) We note, however,
that the constraints from the EGRET measurements do not
change significantly with the EGMF strength as long as the
nucleon flux is comparable
to the $\gamma-$ray flux at the highest energies, as is the
case in Figs.~\ref{fig3} and~\ref{fig4}.
The results of Ref.~\cite{SLBY} differ from
those of Ref.~\cite{ps1} which obtained more stringent
constraints on TD models because of the use of an older
fragmentation function from Ref.~\cite{hill}, and a stronger dependence
on the EGMF because of the use of a weaker EGMF which lead
to a dominance of $\gamma-$rays above $\simeq10^{20}\,$eV.

The energy loss and absorption lengths for UHE nucleons and photons
are short ($\la100$ Mpc). Thus, their predicted UHE fluxes are
independent of cosmological evolution. The $\gamma-$ray flux
below $\simeq10^{11}\,$eV, however, scales as the
total X particle energy release integrated over all redshifts
and increases with decreasing $p$~\cite{sjsb}. For
$m_X=2\times10^{16}\,$GeV,
scenarios with $p<1$ are therefore ruled
out (as can be inferred from Figs.~\ref{fig3} and ~\ref{fig4}), whereas
constant comoving injection models ($p=2$) are well within the
limits.

We now turn to signatures of TD models at UHE.
The full cascade calculations predict
$\gamma-$ray fluxes below 100 EeV that are a factor $\simeq3$
and $\simeq10$ higher than those obtained
using the CEL or absorption approximation often used in the
literature, in the case of strong and weak URB,
respectively. Again, this shows the importance
of non-leading particles in the development of unsaturated EM
cascades at energies below $\sim10^{22}\,$eV.
Our numerical simulations give a $\gamma$/CR flux ratio at
$10^{19}\,$eV of $\simeq0.1$. The experimental exposure
required to detect a $\gamma-$ray flux at that level is
$\simeq4\times10^{19}\,{\rm cm^2}\,{\rm sec}\,{\rm sr}$, about a
factor 10 smaller than the current total experimental exposure.
These exposures are well within reach of the
Pierre Auger Cosmic Ray Observatories~\cite{auger}, which may be able to
detect a neutral CR component down to a level of 1\% of the total
flux. In contrast, if the EGMF exceeds $\sim 10^{-11}\,$G, then UHE
cascading is inhibited, resulting in a lower UHE
$\gamma-$ray spectrum. In the $10^{-9}$ G scenario of Fig.~\ref{fig4},
the $\gamma$/CR flux ratio at $10^{19}\,$eV is $0.02$,
significantly lower than for no EGMF. 

It is clear from the above discussions that the predicted particle fluxes
in the TD scenario are currently uncertain to a large extent due to 
particle physics uncertainties (e.g., mass and decay modes of the X
particles, the quark fragmentation function, the nucleon fraction $f_N$,
and so on) as well as astrophysical uncertainties (e.g., strengths of the
radio and infrared backgrounds, extragalactic magnetic fields, etc.). 
More details on the dependence of the predicted UHE particle spectra and
composition on these particle physics and astrophysical
uncertainties are contained in Ref.~\cite{SLBY}. A detailed
study of the uncertainties involved in the propagation of
UHE nucleons, $\gamma-$rays, and neutrinos is currently
underway~\cite{kkss}.

We stress here that there are viable TD scenarios which
predict nucleon fluxes that are comparable to or even higher than
the $\gamma-$ray flux at all energies, even though $\gamma-$rays
dominate at production.
This occurs, e.g., in the case of high URB
and/or for a strong EGMF, and a nucleon fragmentation fraction of
$\simeq10\%$; see, for example, Fig.~\ref{fig4}. Some of these TD 
scenarios would therefore remain viable even if EHECR induced EAS
should be proven inconsistent with photon primaries (see,
e.g., Ref.~\cite{gamma}). This is in contrast to scenarios with
decaying massive dark matter in the Galactic halo which,
due to the lack of absorption, predict compositions directly
given by the fragmentation function, i.e. domination by
$\gamma-$rays.

The normalization procedure to the EHECR flux described above
imposes the constraint $Q^0_{\rm EHECR}\la10^{-22}\,{\rm eV}\,{\rm
cm}^{-3}\,{\rm sec}^{-1}$ within a factor of a
few~\cite{ps1,SLBY,slsc} for the total energy release rate $Q_0$
from TDs at the current epoch.
In most TD models, because of the unknown values of the
parameters involved, it is currently not
possible to calculate the exact value of $Q_0$ from first principles,
although it has been shown that the required values of $Q_0$ (in order to
explain the EHECR flux) mentioned above are quite possible for
certain kinds of TDs. Some cosmic
string simulations and the necklace scenario suggest that
defects may lose most of
their energy in the form of X particles and estimates of this
rate have been given~\cite{vincent,BV}. If that is the case, the
constraint on $Q^0_{\rm EHECR}$ translates via Eq.~(\ref{dotnx})
into a limit on the symmetry
breaking scale $\eta$ and hence on the mass $m_X$ of the X particle: 
$\eta\sim m_X\la10^{13}\,$GeV~\cite{wmgb}. Independently 
of whether or not this scenario explains EHECR, the EGRET measurement
of the diffuse GeV $\gamma-$ray background leads to a similar bound,
$Q^0_{\rm EM}\la2.2\times10^{-23}\,h
(3p-1)\,{\rm eV}\,{\rm cm}^{-3}\,{\rm sec}^{-1}$, which leaves
the bound on $\eta$ and $m_X$ practically unchanged.
Furthermore, constraints from limits on CMB distortions and light
element abundances from $^4$He-photodisintegration are
comparable to the bound from the directly observed
diffuse GeV $\gamma$-rays~\cite{sjsb}. That these crude
normalizations lead to values of $\eta$ in the right range
suggests that defect models require less fine tuning than
decay rates in scenarios of metastable massive dark matter.

\subsection{Results: Neutrino Fluxes}
As discussed in Sect.~4.1, in TD scenarios most of the energy is
released in the form of EM particles and neutrinos. If the X
particles decay into a quark and a lepton, the quark hadronizes
mostly into pions and the ratio of energy release into the
neutrino versus EM channel is $r\simeq0.3$.

\begin{figure}[htb]
\centerline{\hbox{\psfig{figure=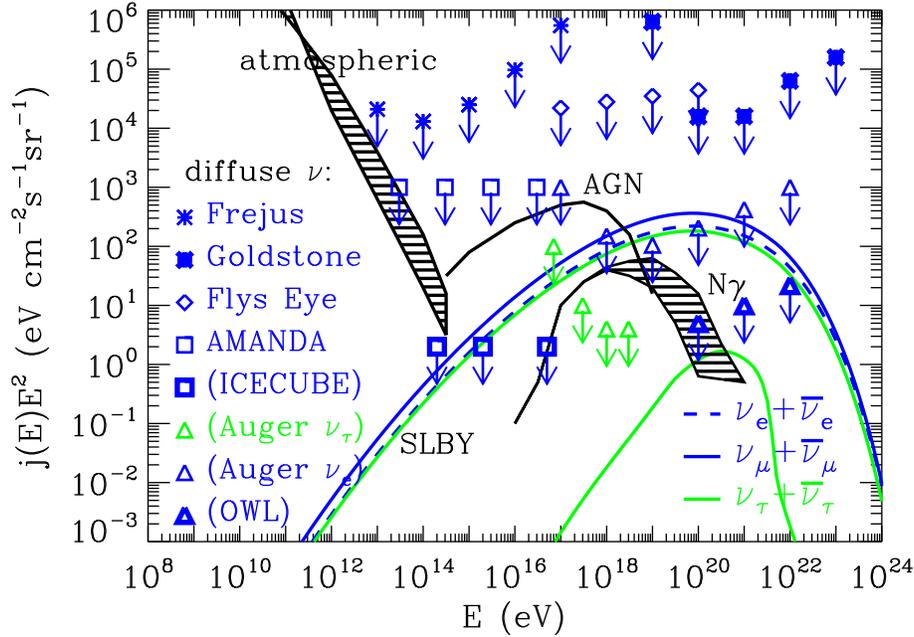,height=0.4\textheight}}}
\caption[...]{\footnotesize Neutrino fluxes predicted by the TD model from
fig.~\ref{fig3}, marked by ``SLBY98''~\cite{SLBY}. Furthermore
shown are experimental neutrino flux limits from the Frejus underground
detector~\cite{frejus}, the Fly's Eye~\cite{baltrusaitis}, the
Goldstone radio telescope~\cite{gln}, and the
Antarctic Muon and Neutrino Detector Array (AMANDA) neutrino
telescope~\cite{amanda}, as well as projected neutrino flux
sensitivities of ICECUBE, the planned kilometer scale extension
of AMANDA~\cite{icecube}, the Pierre Auger Project~\cite{auger-neut}
(for electron and tau neutrinos separately) and the proposed
space based OWL~\cite{owl} concept. For comparison also shown are
the atmospheric neutrino background~\cite{lipari} (hatched region
marked ``atmospheric''),
and neutrino flux predictions for a model of AGN optically thick to
nucleons~\cite{protheroe2}
(``AGN''), and for UHECR interactions with the CMB~\cite{pj}
(``$N\gamma$'', dashed range indicating typical uncertainties
for moderate source evolution).
The top-down fluxes are shown for electron-, muon, and tau-neutrinos
separately, assuming no (lower $\nu_\tau$-curve) and maximal
$\nu_\mu-\nu_\tau$ mixing (upper $\nu_\tau$-curve, which would then
equal the $\nu_\mu$-flux), respectively.}
\label{fig5}
\end{figure}

Fig.~\ref{fig5} shows predictions of the total neutrino
flux for the same TD model on which
Fig.~\ref{fig3} is based. In the absence of neutrino
oscillations the electron neutrino and
anti-neutrino fluxes are about a factor of 2
smaller than the muon neutrino and anti-neutrino fluxes,
whereas the $\tau-$neutrino flux is in general negligible.
In contrast, if the
interpretation of the atmospheric neutrino deficit in terms
of nearly maximal mixing of muon and $\tau-$neutrinos proves
correct, the muon neutrino fluxes shown in Fig.~\ref{fig5} would
be maximally mixed with the $\tau-$neutrino fluxes. To put the TD
component of the neutrino flux in perspective with contributions
from other sources, Fig.~\ref{fig5} also shows the atmospheric
neutrino flux~\cite{lipari}, a typical prediction for the diffuse flux from
photon optically thick proton blazars~\cite{protheroe2} that are not
subject to the Waxman Bahcall bound and were normalized to recent
estimates of the blazar contribution to the diffuse $\gamma-$ray
background~\cite{muk-chiang}, and the flux range expected for
``cosmogenic'' neutrinos created as secondaries from the decay
of charged pions produced by UHE nucleons~\cite{pj}. The TD flux
component clearly dominates above $\sim10^{19}\,$eV.

In order to translate neutrino fluxes into event rates,
one has to fold in the interaction cross sections with
matter. At UHEs these
cross sections are not directly accessible to laboratory
measurements. Resulting uncertainties therefore
translate directly to bounds on neutrino fluxes derived from,
for example, the non-detection of UHE muons produced in charged-current
interactions. In the following, we will assume the estimate
Eq.~(\ref{cross})
based on the Standard Model
for the charged-current muon-neutrino-nucleon
cross section $\sigma_{\nu N}$ if not indicated otherwise.

For an (energy dependent) ice or water equivalent acceptance
$A(E)$ (in units of volume times solid angle), one can obtain an
approximate expected rate of UHE muons produced by neutrinos
with energy $>E$, $R(E)$, by
multiplying $A(E)\sigma_{\nu N}(E)n_{\rm H_2O}$ (where 
$n_{\rm H_2O}$ is the nucleon density in water) with the
integral muon neutrino flux $\simeq Ej_{\nu_\mu}$. This can be used to
derive upper limits on diffuse neutrino fluxes from a
non-detection of muon induced events. Fig.~\ref{fig5} shows bounds
obtained from several experiments: The Frejus experiment
derived upper bounds for $E\ga10^{12}\,$eV from
their non-detection of almost horizontal muons with an energy
loss inside the detector of more than $140\,$MeV per radiation
length~\cite{frejus}. The EAS-TOP collaboration published
two limits from horizontal showers, one in the regime
$10^{14}-10^{15}\,$eV, where non-resonant neutrino-nucleon
processes dominate, and one at the Glashow resonance which
actually only applies to $\bar\nu_e$~\cite{eastop2}. The Fly's
Eye experiment derived
upper bounds for the energy range between $\sim10^{17}\,$eV and
$\sim10^{20}\,$eV~\cite{baltrusaitis} from the non-observation
of deeply penetrating particles.
The AKENO group has published an upper
bound on the rate of near-horizontal, muon-poor air
showers~\cite{nagano}. Horizontal
air showers created by electrons, muons or tau leptons that are in turn
produced by charged-current reactions of electron, muon or tau
neutrinos within the atmosphere have recently also been pointed out
as an important method to constrain or measure UHE neutrino
fluxes~\cite{auger-neut} with next generation detectors.

The $p=0$ TD model BHS0 from the early work of Ref.~\cite{BHS}
is not only ruled out by the constraints from Sect.~4.3,
but also by some of the
experimental limits on the UHE neutrino flux, as can be seen in
Fig.~\ref{fig5}. Further, although both the BHS1 and the SLBY98 models 
correspond to $p=1$, the UHE neutrino flux above $\simeq10^{20}\,$eV in
the latter is almost two orders of magnitude smaller
than in the former. The main reason for this is the different
flux normalization adopted in the two papers: First, the BHS1 model 
was obtained by normalizing the predicted {\it proton} flux to the
observed UHECR flux at $\simeq 4\times10^{19}\,$eV, whereas 
in the SLBY98 model the actually ``visible'' sum of the nucleon and
$\gamma-$ray fluxes was normalized in an optimal way. Second, the BHS1
assumed a nucleon fraction about a factor
3 smaller~\cite{BHS}. Third, the BHS1 scenario used an older fragmentation
function from Ref.~\cite{hill} which has more power at larger energies.
Clearly, the SLBY98 model is not only consistent with the constraints
discussed in Sect.~4.3, but also with all existing neutrino
flux limits within 2-3 orders of magnitude.

What, then, are the prospects of detecting UHE neutrino fluxes
predicted by TD models? In a $1\,{\rm km}^3\,2\pi\,$sr size
detector, the SLBY98 scenario from Fig.~\ref{fig5},
for example, predicts a muon-neutrino event rate
of $\simeq0.15\,{\rm yr}^{-1}$, and an electron neutrino event rate
of $\simeq0.089\,{\rm yr}^{-1}$ above $10^{19}\,$eV, where
``backgrounds'' from conventional sources should be negligible.
Further, the muon-neutrino event rate above 1 PeV should be
$\simeq1.2\,{\rm yr}^{-1}$, which could be interesting if
conventional sources produce neutrinos at a much smaller
flux level. Of course, above $\simeq100\,$TeV, instruments
using ice or water as detector medium, have to look at downward
going muon and electron events due to neutrino absorption in
the Earth. However, $\tau-$neutrinos obliterate this Earth
shadowing effect due to their regeneration from $\tau$
decays~\cite{halzen-saltzberg}. The presence of $\tau-$neutrinos,
for example, due to mixing with muon neutrinos, as suggested
by recent experimental results from Super-Kamiokande,
can therefore lead to an increased upward going event
rate~\cite{mannheim3}. For recent compilations of UHE neutrino flux
predictions from astrophysical and TD sources see
Refs.~\cite{owl1,neutflux} and references therein.

For detectors based on the fluorescence technique such as the
HiRes~\cite{hires} and the Telescope Array~\cite{tel_array}
(see Sect.~2), the sensitivity to UHE neutrinos is often
expressed in terms of an effective aperture $a(E)$ which is
related to $A(E)$ by $a(E)=A(E)\sigma_{\nu N}(E)n_{\rm
H_2O}$. For the cross section of Eq.~(\ref{cross}), the
apertures given in Ref.~\cite{hires} for the HiRes correspond to
$A(E)\simeq3\,{\rm km}^3\times2\pi\,{\rm sr}$ for
$E\ga10^{19}\,$eV for muon neutrinos. The expected acceptance
of the ground array component of the Pierre Auger
project for horizontal
UHE neutrino induced events is $A(10^{19}\,{\rm eV})\simeq
20\,{\rm km}^3\,{\rm sr}$ and $A(10^{23}\,{\rm eV})\simeq
200\,{\rm km}^3\,{\rm sr}$~\cite{auger-neut}, with a duty cycle close to
100\%. We conclude that detection of neutrino
fluxes predicted by scenarios such as the SLBY98 scenario shown
in Fig.~\ref{fig5} requires running a detector of acceptance
$\ga10\,{\rm km}^3\times2\pi\,{\rm sr}$ over a period of a few
years. Apart from optical detection in air, water, or ice, other
methods such as acoustical and radio detection~\cite{ghs}
(see, e.g., the RICE project~\cite{rice} for the latter) or even
detection from space~\cite{owl,owl1,euso,airwatch} appear to be interesting
possibilities for detection concepts operating at such scales
(see Sect.~2). For example, the space based OWL/AirWatch satellite
concept would have an
aperture of $\simeq3\times10^6\,{\rm km}^2\,{\rm sr}$
in the atmosphere, corresponding to $A(E)\simeq6\times10^4
\,{\rm km}^3\,{\rm sr}$ for $E\ga10^{20}\,$eV,
with a duty cycle of $\simeq0.08$~\cite{owl,owl1}.
The backgrounds seem to be
in general negligible~\cite{ydjs,price}. As indicated by the
numbers above and by the projected sensitivities shown in
Fig.~\ref{fig5}, the Pierre Auger Project and especially the
space based AirWatch type projects should be capable of
detecting typical TD neutrino
fluxes. This applies to any detector of acceptance
$\ga100\,{\rm km}^3\,{\rm sr}$. Furthermore, a 100 day
search with a radio telescope of the NASA Goldstone type
for pulsed radio emission from cascades induced by neutrinos
or cosmic rays in the lunar regolith could reach
a sensitivity comparable or better to the Pierre Auger
sensitivity above $\sim10^{19}\,$eV~\cite{gln}.

A more model independent estimate~\cite{slsc} for the average
event rate $R(E)$ can be made if the underlying
scenario is consistent with observational nucleon and
$\gamma-$ray fluxes and the bulk of the energy is released above
the PP threshold on the CMB. Let us assume that the
ratio of energy injected into the neutrino versus EM channel is a
constant $r$. As discussed in Sect.~4.3, cascading effectively reprocesses
most of the injected EM energy into low
energy photons whose spectrum peaks at $\simeq10\,$GeV~\cite{ahacoppi}.
Since the ratio $r$ remains roughly unchanged during
propagation, the height of the
corresponding peak in the neutrino spectrum should
roughly be $r$ times the height of the low-energy
$\gamma-$ray peak, i.e., we have the condition
$\max_E\left[E^2j_{\nu_\mu}(E)\right]\simeq
r\max_E\left[E^2j_\gamma(E)\right].$ Imposing the observational
upper limit on the diffuse $\gamma-$ray flux around $10\,$GeV
shown in Fig.~\ref{fig5}, $\max_E\left[E^2j_{\nu_\mu}(E)\right]\la
2\times10^3 r \,{\rm eV}{\rm cm}^{-2}{\rm sec}^{-1}{\rm
sr}^{-1}$, then bounds the average diffuse neutrino rate above
PP threshold on the CMB, giving 
\begin{equation}
  R(E)\la0.34\,r\left[{A(E)\over1\,{\rm
  km}^3\times2\pi\,{\rm sr}}\right]
  \,\left({E\over10^{19}\,{\rm eV}}\right)^{-0.6}\,{\rm
  yr}^{-1}\quad(E\ga10^{15}\,{\rm eV})\,.\label{r2}
\end{equation}
For $r\la20(E/10^{19}\,{\rm eV})^{0.1}$ this bound is consistent
with the flux bounds shown in Fig.~\ref{fig5} that are dominated
by the Fly's Eye constraint at UHE. We stress again that TD
models are not subject to the Waxman Bahcall bound because
the nucleons produced are considerably less abundant than
and are not the primaries of produced $\gamma-$rays and
neutrinos.

\begin{figure}[htb]
\centerline{\hbox{\psfig{figure=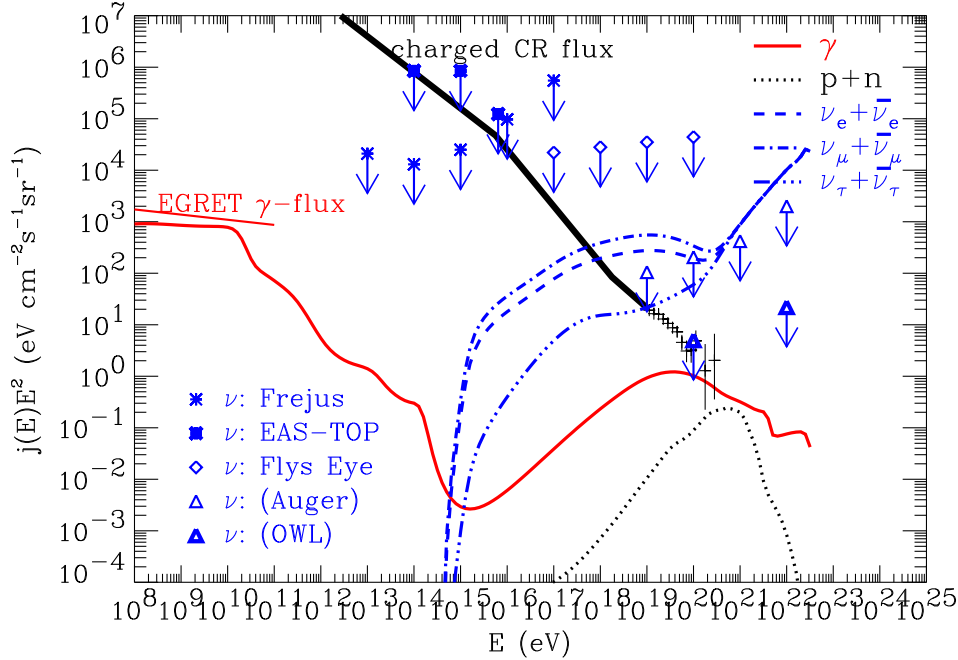,height=0.4\textheight}}}
\caption[...]{\footnotesize Flux predictions for a TD model characterized
by $p=1$, $m_X=10^{14}\,$GeV, with X particles exclusively
decaying into neutrino-antineutrino pairs of all flavors
(with equal branching ratio), assuming a neutrino mass $m_\nu=1\,$eV. For
neutrino clustering, an overdensity of $\simeq30$ over
a scale of $l_\nu\simeq5\,$Mpc was assumed. The calculation
assumed the strongest
URB version from Ref.~\cite{PB} and an EGMF $\ll10^{-11}\,$G.
The line key is as in Figs.~\ref{fig3} and~\ref{fig5}.
\label{fig6}}
\end{figure}

In typical TD models such as the one discussed above where
primary neutrinos are produced by pion decay,
$r\simeq0.3$. However, in TD scenarios with $r\gg1$ neutrino
fluxes are only limited by the condition that the {\it secondary}
$\gamma-$ray flux produced by neutrino interactions with the
relic neutrino background be below
the experimental limits. An example for such a scenario is given
by X particles exclusively decaying into neutrinos
(although this is not very likely in most particle physics
models, but see Ref.~\cite{SLBY} and Fig.~\ref{fig6} for a scenario
involving topological defects and Ref.~\cite{gk2} for a scenario
involving decaying superheavy relic particles, both of which explain the
observed EHECR events as secondaries of neutrinos interacting
with the primordial neutrino background). Such scenarios predict
appreciable event rates above $\sim10^{19}\,$eV in a km$^3$
scale detector, but require unrealistically strong clustering
of relic neutrinos (a homogeneous relic neutrino overdensity
would make the EGRET constraint only more severe
because neutrino interactions beyond $\sim50\,$Mpc
contribute to the GeV $\gamma-$ray background but not
to the UHECR flux). A detection would thus open the exciting
possibility to establish an experimental lower limit on $r$.
Being based solely on energy conservation,
Eq.~(\ref{r2}) holds regardless of whether or not the underlying
TD mechanism explains the observed EHECR events.

The transient neutrino event rate could be much higher than
Eq.~(\ref{r2}) in the direction to discrete sources which emit
particles in bursts. Corresponding pulses in the EHE
nucleon and $\gamma-$ray fluxes would only occur for sources
nearer than $\simeq100\,$Mpc and, in case of protons, would be
delayed and dispersed by deflection in Galactic and
extragalactic magnetic fields~\cite{wm,lsos}. The recent observation
of a possible clustering of CRs above $\simeq4\times10^{19}\,$eV by the
AGASA experiment~\cite{haya2} might suggest
sources which burst on a time scale $t_b\ll1\,$yr.
A burst fluence of $\simeq r\left[A(E)/1\,{\rm km}^3\times2\pi\,{\rm
sr}\right](E/10^{19}\,{\rm eV})^{-0.6}$ neutrino induced events
within a time $t_b$ could then be expected. Associated pulses
could also be observable in the ${\rm GeV}-{\rm TeV}$
$\gamma-$ray flux if the EGMF is smaller than
$\simeq10^{-15}\,$G in a significant fraction of extragalactic
space~\cite{wc}.

In contrast to roughly homogeneous sources and/or mechanisms
with branching ratios $r\gg1$, in scenarios involving
clustered sources such as metastable superheavy relic
particles decaying with $r\sim1$, the neutrino flux is
comparable to (not significantly larger than) the UHE photon
plus nucleon fluxes.
This can be understood because the neutrino flux is dominated
by the extragalactic contribution which scales with the
extragalactic nucleon and $\gamma-$ray contribution in exactly
the same way as in the unclustered case, whereas the extragalactic
contribution to the ``visible'' flux to be normalized to the
UHECR data is much smaller in the clustered case.
The resulting neutrino fluxes would be hardly detectable even
with next generation experiments.

\section{UHE Cosmic Rays and Cosmological Large Scale Magnetic Fields}

\subsection{Deflection and Delay of Charged Hadrons}
Whereas for UHE electrons the dominant influence of large scale
magnetic fields is
synchrotron loss rather than deflection, for charged hadrons the
opposite is the case. A relativistic particle of charge $qe$ and energy $E$
has a gyroradius $r_g\simeq E/(qeB_\perp)$ where $B_\perp$ is the
field component perpendicular to the particle momentum. If this
field is constant over a distance $d$, this leads to a
deflection angle
\begin{equation}
  \theta(E,d)\simeq\frac{d}{r_g}\simeq0.52^\circ q
  \left(\frac{E}{10^{20}\,{\rm eV}}\right)^{-1}
  \left(\frac{d}{1\,{\rm Mpc}}\right)
  \left(\frac{B_\perp}{10^{-9}\,{\rm G}}\right)
  \,.\label{gyro}
\end{equation}

Magnetic fields beyond the Galactic disk are poorly known and
include a possible extended field in the halo of our Galaxy and
a large scale EGMF. In both cases, the magnetic
field is often characterized by an r.m.s. strength $B$ and a
correlation length $l_c$, i.e. it is assumed that
its power spectrum has a cut-off in wavenumber space at
$k=2\pi/l_c$ and in real space it is smooth on scales below
$l_c$. If we neglect energy loss processes for the moment, then
the r.m.s. deflection angle over a distance $d$ in such a field
is $\theta(E,d)\simeq(2dl_c/9)^{1/2}/r_g$, or
\begin{equation}
  \theta(E,d)\simeq0.8^\circ\,
  q\left(\frac{E}{10^{20}\,{\rm eV}}\right)^{-1}
  \left(\frac{d}{10\,{\rm Mpc}}\right)^{1/2}
  \left(\frac{l_c}{1\,{\rm Mpc}}\right)^{1/2}
  \left(\frac{B}{10^{-9}\,{\rm G}}\right)\,,\label{deflec}
\end{equation}
for $d\ga l_c$, where the numerical prefactors were calculated
from the analytical treatment in Ref.~\cite{wm}. There it was
also pointed
out that there are two different limits to distinguish: For
$d\theta(E,d)\ll l_c$, particles of all energies 
``see'' the same magnetic field realization during their
propagation from a discrete source to the observer. In this
case, Eq.~(\ref{deflec}) gives the typical coherent deflection
from the line-of-sight source direction, and the spread in arrival
directions of particles of different energies 
is much smaller. In contrast, for $d\theta(E,d)\gg l_c$, the
image of the source is washed out over a typical angular extent
again given by Eq.~(\ref{deflec}), but in this case it is 
centered on the true source direction. If $d\theta(E,d)\simeq
l_c$, the source may even have several images, similar to the
case of gravitational lensing. Therefore, observing
images of UHECR sources and identifying counterparts in other
wavelengths would allow one to distinguish these limits and thus
obtain information on cosmic magnetic fields. If $d$ is
comparable to or larger than the interaction length for stochastic
energy loss due to photo-pion production or photodisintegration,
the spread in deflection angles is always comparable to the
average deflection angle.

Deflection also implies an average time delay of
$\tau(E,d)\simeq d\theta(E,d)^2/4$, or
\begin{equation}
  \tau(E,d)\simeq1.5\times10^3\,q^2
  \left(\frac{E}{10^{20}\,{\rm eV}}\right)^{-2}
  \left(\frac{d}{10\,{\rm Mpc}}\right)^{2}
  \left(\frac{l_c}{1\,{\rm Mpc}}\right)
  \left(\frac{B}{10^{-9}\,{\rm G}}\right)^2\,{\rm yr}
  \label{delay}
\end{equation}
relative to rectilinear propagation with the speed of light. It was 
pointed out in Ref.~\cite{mw} that, as a consequence, the observed
UHECR spectrum of a bursting source at a given time can be
different from its long-time average and would typically peak
around an energy $E_0$, given by equating $\tau(E,d)$ with the time of
observation relative to the time of arrival for vanishing time
delay. Higher energy particles would have passed the observer
already, whereas lower energy particles would not have arrived
yet. Similarly to the behavior of deflection angles, the width
of the spectrum around $E_0$ would be much smaller than $E_0$ if
both $d$ is smaller than the interaction length for stochastic
energy loss and $d\theta(E,d)\ll l_c$. In all other cases the
width would be comparable to $E_0$.

Constraints on magnetic fields from deflection and time delay
cannot be studied separately from the characteristics of the
``probes'', namely the UHECR sources, at least as long as their
nature is unknown. An approach to the general case is discussed
in Sect.~5.3.

\subsection{Constraints on EHECR Source Locations}
As pointed out in Sect.~1, nucleons, nuclei, and $\gamma-$rays
above a few $10^{19}\,$eV
cannot have originated much further away than
$\simeq50\,$Mpc. Together with Eq.~(\ref{deflec}) this implies
that above a few $10^{19}\,$eV the arrival direction of such
particles should in general point back to their source within a few
degrees~\cite{ssb}. This argument is often made in the literature
and follows from the Faraday rotation bound on the EGMF and a
possible extended field in the halo of our Galaxy, which in its
historical form reads
$Bl_c^{1/2}\la10^{-9}\,{\rm G}\,{\rm Mpc}^{1/2}$~\cite{mag-rev},
as well as from the known strength and
scale height of the field in the disk of our Galaxy,
$B_g\simeq3\times10^{-6}\,$G, $l_g\la1\,$kpc. Furthermore, the
deflection in the disk of our Galaxy can be corrected for in
order to reconstruct the extragalactic arrival direction: Maps
of such corrections as a function of arrival direction have been
calculated in Refs.~\cite{gal-deflec} for plausible models of
the Galactic magnetic field. The deflection of UHECR trajectories in
the Galactic magnetic field may, however, also give rise to several other 
important effects~\cite{hmr} such as (de)magnification of the UHECR
fluxes due to the magnetic lensing effect mentioned in the previous section 
(which can modify the UHECR spectrum from individual sources), 
formation of multiple images of a source, and apparent ``blindness'' of
the Earth towards certain regions of the sky with regard to UHECRs.
These effects may in turn have important implications for UHECR source
locations. In fact, it was recently claimed~\cite{ambs}
that, assuming a certain model of the magnetic fields
in the galactic winds, the highest energy cosmic ray events
could all have originated in the Virgo cluster or specifically
in the radio galaxy M87. However, as was subsequently 
pointed out in Ref.~\cite{bls}, this galactic wind model
leads to focusing of all positively charged highest
energy particles to the North galactic pole and, consequently,
this can not be interpreted as evidence for a point
source situated close to the North galactic pole.

However, important modifications of the Faraday rotation bound
on the EGMF have recently been discussed in the literature:
The average electron density which enters estimates of the EGMF
from rotation measures, can now be more reliably estimated from
the baryon density $\Omega_bh^2\simeq0.02$, whereas in the original
bound the closure density was used. Assuming an unstructured
Universe and $\Omega_0=1$ results in the much weaker bound~\cite{bbo}
\begin{equation}
  B\la3\times10^{-7}\left(\frac{\Omega_bh^2}{0.02}\right)^{-1}
  \left(\frac{h}{0.65}\right)
  \left(\frac{l_c}{{\rm Mpc}}\right)^{-1/2}
  \,{\rm G}\,,\label{newFhom}
\end{equation}
which suggests much stronger deflection. However, taking into
account the large scale structure of the Universe in the form
of voids, sheets, filaments etc., and assuming flux freezing
of the magnetic fields whose strength then approximately scales
with the 2/3 power of the local density, leads to more stringent bounds:
Using the Lyman $\alpha$ forest to model the density distribution
yields~\cite{bbo}
\begin{equation}
  B\la10^{-9}-10^{-8}\,{\rm G}\label{newF}
\end{equation}
for the large scale EGMF for coherence scales between the Hubble
scale and 1 Mpc. This estimate is closer to the original Faraday
rotation limit. However, in this scenario the maximal fields
in the sheets and voids can be as high as a
$\mu\,$G~\cite{rkb,bbo}.

Therefore, according to Eq.~(\ref{deflec}) and~(\ref{newF}),
deflection of UHECR nucleons is still expected to be on
the degree scale if the local large scale structure around the
Earth is not strongly magnetized. However,
rather strong deflection can occur if the Supergalactic Plane
is strongly magnetized, for particles originating in
nearby galaxy clusters where magnetic
fields can be as high as $10^{-6}\,$G~\cite{mag-rev}
(see Sect.~5.3 below) and/or for
heavy nuclei such as iron~\cite{elb-som}. In this case, magnetic
lensing in the EGMF can also play an important role in
determining UHECR source locations~\cite{slb,lsb}.

\subsection{Angle-Energy-Time Images of UHECR Sources}

\subsection*{Small Deflection}

For small deflection angles and if photo-pion
production is important, one has to resort to numerical Monte
Carlo simulations in 3 dimensions. Such simulations have been
performed in Ref.~\cite{tph} for the case $d\theta(E,d)\gg l_c$
and in Refs.~\cite{lsos,slo,sl} for the general case.

\begin{figure}[htb]
\centerline{\hbox{\psfig{figure=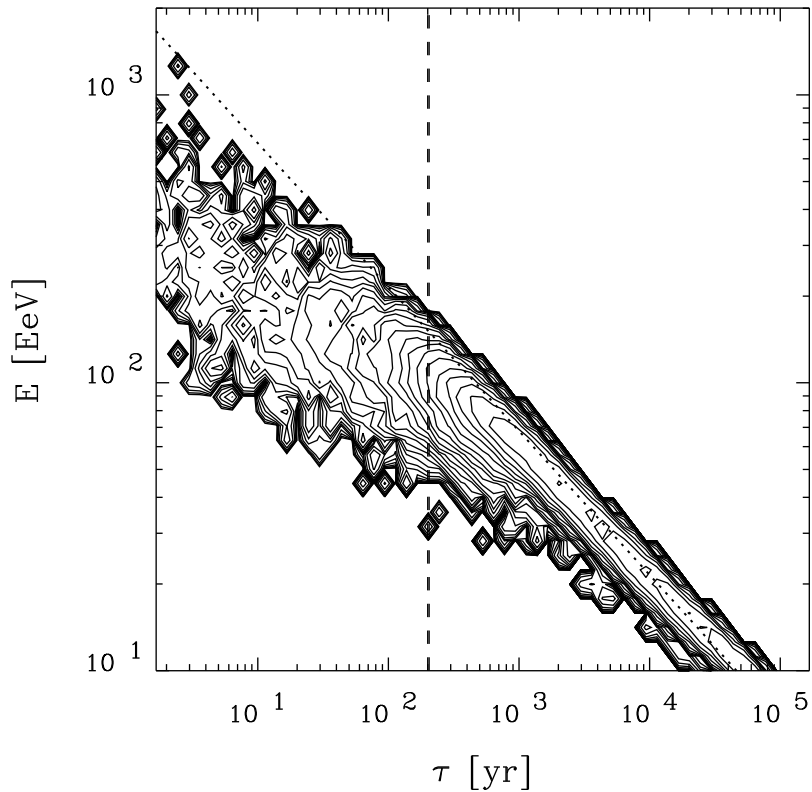,height=0.5\textheight}}}
\caption[...]{\footnotesize Contour plot of the UHECR image of a
bursting source at $d=30\,$Mpc, projected onto the time-energy plane, with
$B=2\times10^{-10}\,$G, $l_c=1\,$Mpc, from Ref.~\cite{lsos}.
The contours decrease in steps of 0.2 in the logarithm to base 10.
The dotted line indicates the energy-time delay correlation 
$\tau(E,d)\propto E^{-2}$ as would be obtained in the absence of 
pion production losses. Clearly, $d\theta(E,d)\ll l_c$ in this
example, since for $E<4\times10^{19}\,$eV, the width of the
energy distribution at any given time is much smaller than the
average (see Sect.~5.1). The dashed lines, which are not resolved 
here, indicate the location (arbitrarily chosen) of the
observational window, of length $T_{obs}=5\,$yr.}
\label{fig7}
\end{figure}

In Refs.~\cite{lsos,slo,sl} the Monte Carlo simulations were
performed in the following way:
The magnetic field was represented as a Gaussian random field
with zero mean and a power spectrum
with $\left\langle B^2(k)\right\rangle\propto
k^{n_H}$ for $k<k_c$ and $\left\langle B^2(k)\right\rangle=0$
otherwise, where $k_c=2\pi/l_c$ characterizes the numerical
cut-off scale and the r.m.s. strength is
$B^2=\int_0^\infty\,dk\,k^2\left\langle B^2(k)
\right\rangle$. The field is then calculated on a
grid in real space via Fourier transformation.
For a given magnetic field realization and source, nucleons with
a uniform logarithmic distribution of injection energies are
propagated between two given points (source and observer) on the
grid. This is done by solving the equations of motion in the
magnetic field interpolated between the grid points, and
subjecting nucleons to stochastic production of pions and (in
case of protons) continuous loss of energy due to PP.
Upon arrival, injection and detection energy, and time
and direction of arrival are recorded. From many (typically
40000) propagated particles, a histogram of average number of
particles detected as a function of time and energy of
arrival is constructed for any given injection spectrum by
weighting the injection energies correspondingly. This histogram
can be scaled to any desired total fluence at the detector and,
by convolution in time, can be constructed for arbitrary
emission time scales of the source. An example for the
distribution of arrival times and energies of UHECRs from a
bursting source is given in Fig.~\ref{fig7}.

We adopt the following notation for the parameters: $\tau_{100}$ 
denotes the time delay due to magnetic deflection at $E=100\,$EeV
and is given by Eq.~(\ref{delay}) in terms of the magnetic field
parameters;
$T_S$ denotes the emission time scale of the source; $T_S\ll1$yr 
corresponds to a burst, and $T_S\gg1$yr (roughly speaking) to a 
continuous source; $\gamma$ is the differential index of the 
injection energy spectrum; $N_0$ denotes the fluence of the 
source with respect to the detector, {\it i.e.}, the total number 
of particles that the detector would detect from the source on 
an infinite time scale; finally, ${\cal L}$ is the likelihood
function of the above parameters.

\begin{figure}[htb]
\centerline{\hbox{\psfig{figure=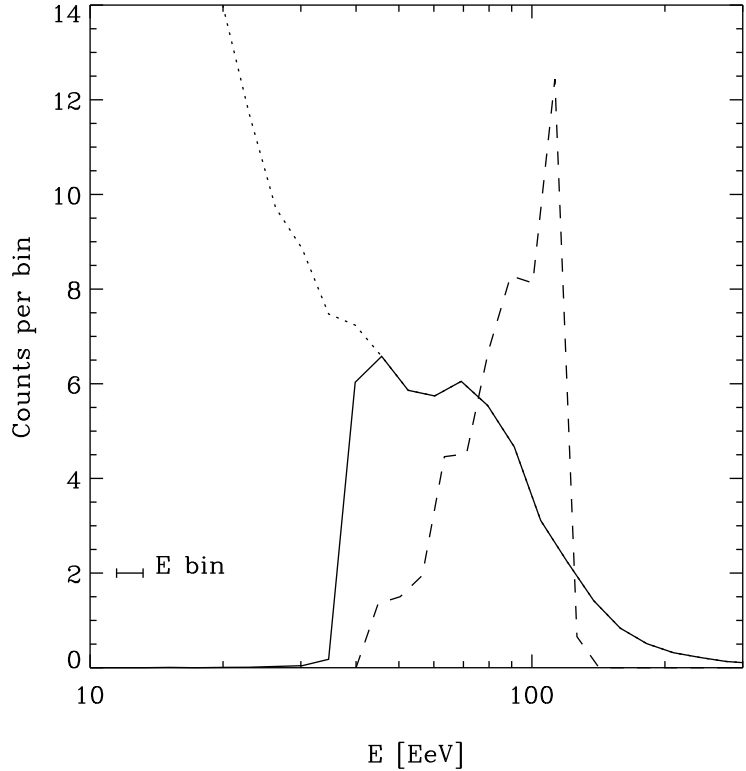,height=0.5\textheight}}}
\caption[...]{\footnotesize Energy spectra for a continuous source (solid 
line), and for a burst (dashed line), from
Ref.~\cite{lsos}. Both spectra are
normalized to a total of 50 particles detected. The parameters 
corresponding to the continuous source case are: $T_S=10^4\,$yr, 
$\tau_{100}=1.3\times10^3\,$yr, and the time of observation is 
$t=9\times10^3\,$yr, relative to rectilinear propagation with the speed
of light. A low energy cutoff results at the energy
$E_S=4\times10^{19}\,$eV where $\tau_{E_S}=t$. The dotted line
shows how the spectrum would continue if $T_S\ll10^4\,$yr. The
case of a bursting source corresponds to a slice of the image in 
the $\tau_E-E$ plane, as indicated in Fig.~\ref{fig7} by dashed lines. 
For both spectra, $d=30\,$Mpc, and $\gamma=2.$.}
\label{fig8}
\end{figure}

By putting windows of width equal to the time
scale of observation over these histograms one obtains expected
distributions of
events in energy and time and direction of arrival
for a given magnetic field realization, source distance and
position, emission time scale, total fluence, and injection
spectrum. Examples of the resulting energy spectrum are shown in
Fig.~\ref{fig8}. By dialing Poisson statistics on such
distributions, one can simulate corresponding observable event
clusters.

Conversely, for any given real or simulated event cluster, one
can construct a likelihood of the observation as a function of 
the time delay, the emission time scale, the differential 
injection spectrum index, the fluence, and the distance. In
order to do so, and to obtain the maximum of the likelihood, 
one constructs histograms for many different parameter
combinations as described above, randomly puts observing time
windows over the histograms, calculates the likelihood function
from the part of the histogram within the window and the cluster
events, and averages over different window locations and magnetic
field realizations.

\begin{figure}[htb]
\centerline{\hbox{\psfig{figure=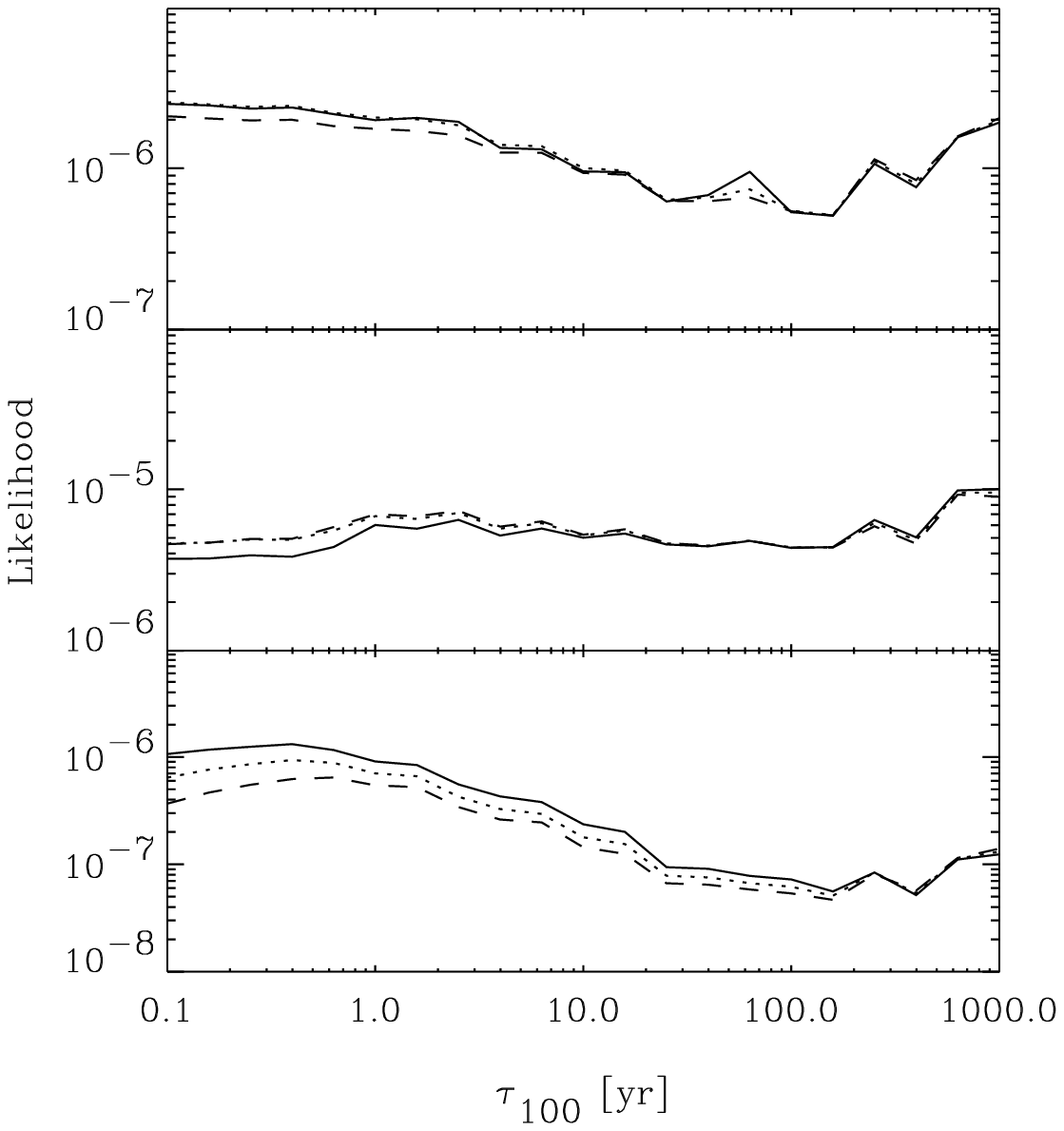,height=0.5\textheight}}}
\caption[...]{\footnotesize The likelihood, ${\cal L}$,
marginalized over $T_S$ and $N_0$ as a function of the average
time delay at $10^{20}\,$eV, $\tau_{100}$, assuming a source
distance $d=30\,$Mpc. The panels are for pair \# 3 through \# 1,
from top to bottom, of the AGASA pairs~\cite{haya2}.
Solid lines are for $\gamma=1.5$, dotted
lines for $\gamma=2.0$, and dashed lines for $\gamma=2.5$.}
\label{fig9}
\end{figure}

In Ref.~\cite{slo} this approach has been applied to and discussed in
detail for the three pairs observed by the AGASA
experiment~\cite{haya2}, under the assumption that all events
within a pair were produced by the same discrete source.
Although the inferred angle between the momenta of the paired
events acquired in the EGMF is several degrees~\cite{medinatanco2},
this is not necessarily evidence against a common source,
given the uncertainties in the Galactic field and the angular
resolution of AGASA which is $\simeq2.5^\circ$. As a result
of the likelihood analysis, these pairs do not seem to follow
a common characteristic; one of them seems to favor a burst, another
one seems to be more consistent with a continuously emitting source.
The current data, therefore, does not allow one to rule out any of
the models of UHECR sources. Furthermore, two of the three pairs are
insensitive to the time delay. However, the pair which contains the
$200\,$EeV event seems to significantly favor a comparatively
small average time delay, $\tau_{100}\la10\,$yr, as can be
seen from the likelihood function marginalized over $T_S$ and $N_0$
(see Fig.~\ref{fig9}). According to Eq.~(\ref{delay}) this
translates into a tentative bound for the r.m.s. magnetic field, namely, 
\begin{equation}
  B\la2\times10^{-11}
  \left(\frac{l_c}{1\,{\rm Mpc}}\right)^{-1/2}
  \left(\frac{d}{30\,{\rm Mpc}}\right)^{-1}\,{\rm G}
  \,,\label{limit}
\end{equation}
which also applies to magnetic fields in the halo of our Galaxy
if $d$ is replaced by the lesser of the source distance and the linear
halo extent. If confirmed by future data, this bound would be at least
two orders of magnitude more restrictive than the best existing
bounds which come from Faraday rotation measurements [see
Eq.~(\ref{newF})] and, for a homogeneous EGMF, from CMB
anisotropies~\cite{bfs}. UHECRs are therefore at least as sensitive
a probe of cosmic magnetic fields as other measures in the range
near existing limits such as the polarization~\cite{lk} and the
small scale anisotropy~\cite{sb} of the CMB.

More generally, confirmation of a clustering of
EHECRs would provide significant information on both the nature
of the sources and on large-scale magnetic
fields~\cite{sslh}. This has been shown quantitatively~\cite{sl} by
applying the hybrid Monte Carlo likelihood analysis discussed
above to simulated clusters of a few tens of events as they
would be expected from next generation experiments~\cite{icrr96} such
as the High Resolution Fly's Eye~\cite{hires}, the Telescope
Array~\cite{tel_array}, and most notably, the Pierre Auger
Project~\cite{auger} (see Sect.~2), provided the clustering recently
suggested by the AGASA experiment~\cite{haya2,agasa2} is real.
The proposed AirWatch type satellite observatory
concepts~\cite{owl,euso,airwatch} might even allow one to
detect clusters of hundreds of such events. 

Five generic situations of UHECR time-energy images
were discussed in Ref.~\cite{sl}, classified according to
the values of the time delay 
$\tau_E$ induced by the magnetic field, the emission timescale of the 
source $T_{\rm S}$, as compared to the lifetime of the experiment.
The likelihood calculated for the simulated clusters in these
cases presents different degeneracies between
different parameters, which complicates the analysis. As an example,
the likelihood is degenerate in the ratios $N_0/T_{\rm S}$, or
$N_0/\Delta\tau_{100}$, where $N_0$ is the total fluence, and
$\Delta\tau_{100}$ is the spread in arrival time; these ratios
represent rates of detection. Another example is given by the
degeneracy between the distance $d$ and the injection energy 
spectrum index $\gamma$. Yet another is the ratio
$(d\tau_E)^{1/2}/l_c$,
that controls the size of the scatter around the mean of the 
$\tau_E-E$ correlation. Therefore, in most general cases, values for 
the different parameters cannot be pinned down, and generally, only 
domains of validity are found. In the following the reconstruction
quality of the main parameters considered is summarized.

The distance to the source can be obtained from the pion 
production signature, above the GZK cut-off, when
the emission timescale of the source dominates over the time delay. 
Since the time delay decreases with increasing energy, the lower the 
energy $E_{\rm C}$, defined by $\tau_{E_{\rm C}}\simeq T_{\rm S}$,
the higher the accuracy on the distance $d$. The error on $d$ is,
in the best case, typically a factor 2, for one cluster of
$\simeq40$ events.
In this case, where the emission timescale dominates over the time 
delay at all observable energies,
information on the magnetic field is only contained in the
angular image, which was not systematically included in the
likelihood analysis of Ref.~\cite{sl} due to computational limits.
Qualitatively, the size of the angular image is 
proportional to $B(dl_c)^{1/2}/E$, whereas the structure
of the image, {\it i.e.}, the number of separate images, is
controlled by the ratio $d^{3/2}B/l_c^{1/2}/E$. Finally, in the
case when the time delay 
dominates over the emission timescale, with a time delay shorter than 
the lifetime of the experiment, one can also estimate the distance 
with reasonable accuracy.

Some sensitivity to the injection spectrum index $\gamma$ exists
whenever events are recorded over a sufficiently broad energy
range. At least if the distance $d$ is known, it is in general
comparatively easy to rule out a hard injection spectrum if the
actual $\gamma\ga2.0$, but much harder to distinguish between
$\gamma=2.0$ and 2.5.

If the lifetime of the experiment is the largest time scale
involved, the strength of the magnetic field can only be obtained
from the time-energy image because the angular image
will not be resolvable. When the time delay 
dominates over the emission timescale, and is, at the same time,
larger than the lifetime of the experiment, only a lower limit 
corresponding to this latter timescale, can be placed on the time 
delay and hence on the strength of the magnetic field. When combined
with the Faraday rotation upper limit Eq.~(\ref{newF}),
this would nonetheless allow one to bracket the r.m.s. magnetic
field strength within a
few orders of magnitude. In this case also, significant information is
contained in the angular image. If the emission time scale is
larger then the delay time, the angular image is obviously the
only source of information on the magnetic field strength.

The coherence length $l_c$ enters in the ratio
$(d\tau_E)^{1/2}/l_c$
that controls the scatter around the mean of the $\tau_E-E$
correlation in the time-energy image. It can therefore be estimated
from the width of this image, provided the emission timescale is
much less than $\tau_E$ (otherwise the correlation would not be seen),
and some prior information on $d$ and $\tau_E$ is available.

\begin{figure}[htb]
\centerline{\hbox{\psfig{figure=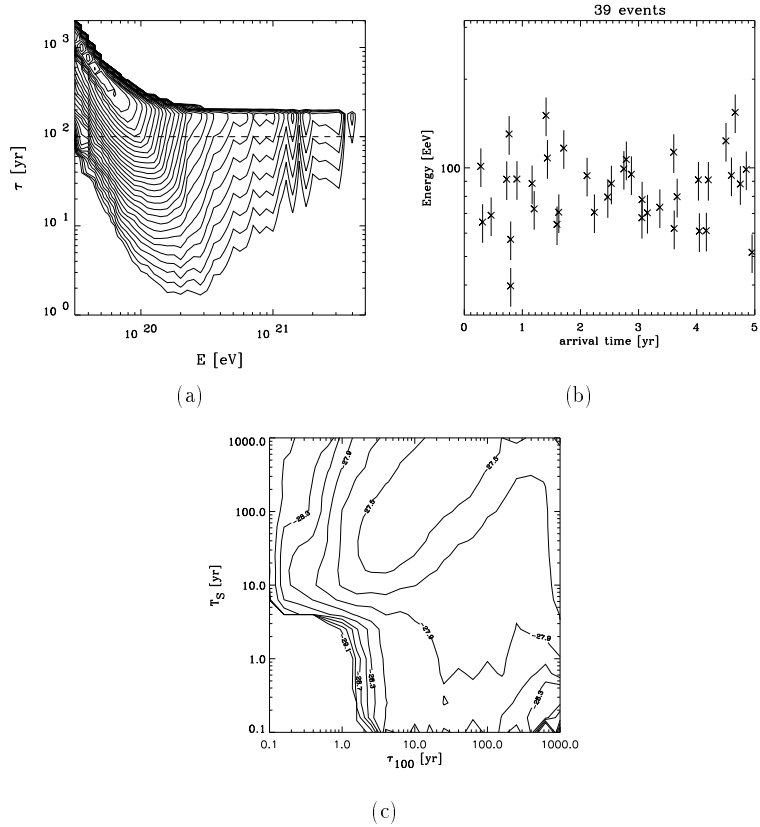,height=0.5\textheight}}}
\caption[...]{\footnotesize (a) Arrival time-energy histogram for
$\gamma=2.0$, $\tau_{100}=50\,$yr, $T_{\rm S}=200\,$yr,
$l_c\simeq1\,$Mpc, $d=50\,$Mpc, corresponding to $B
\simeq3\times10^{-11}\,$G. Contours are in steps of a factor
$10^{0.4}=2.51$; (b) Example of a cluster in the
arrival time-energy plane resulting from the cut indicated in
(a) by the dashed line at $\tau\simeq100\,$yr; (c)
The likelihood function, marginalized over $N_0$
and $\gamma$, for $d=50\,$Mpc, $l_c\simeq\,$Mpc, for the cluster
shown in (b), in the $T_{\rm S}-\tau_{100}$
plane. The contours shown go from the maximum down to about 0.01
of the maximum in steps of a factor $10^{0.2}=1.58$. Note that
the likelihood clearly favors $T_{\rm S}\simeq4\tau_{100}$. For
$\tau_{100}$ large enough to be estimated from the angular
image size, $T_{\rm S}\gg T_{\rm obs}$ can, therefore, be
estimated as well.}
\label{fig10}
\end{figure}

An emission timescale much larger than the experimental lifetime
may be estimated if a lower cut-off in the spectrum is
observable at an energy $E_{\rm C}$, indicating that $T_{\rm
S}\simeq\tau_{E_{\rm C}}$. The latter may, in turn, be estimated
from the angular image size via Eq.~(\ref{delay}), where the
distance can be estimated from the spectrum visible above the
GZK cut-off, as discussed above. An example
of this scenario is shown in Fig.~\ref{fig10}. For angular
resolutions $\Delta\theta$, timescales in the range
\begin{equation}
  3\times10^3\,\left(\frac{\Delta\theta}{1^\circ}\right)^2
  \left(\frac{d}{10\,{\rm Mpc}}\right)\,{\rm yr}
  \la T_{\rm S}\simeq\tau_E\la10^4\cdots10^7\,
  \left(\frac{E}{100\,{\rm EeV}}\right)^{-2}\,{\rm yr}
  \label{tsscale}
\end{equation}
could be probed. The lower limit follows from the requirement
that it should be possible to estimate $\tau_E$ from $\theta_E$,
using Eq.~(\ref{delay}), otherwise only an upper limit on $T_{\rm
S}$, corresponding to this same number, would apply.
The upper bound in Eq.~(\ref{tsscale}) comes from constraints on
maximal time delays in cosmic magnetic fields, such as the Faraday
rotation limit in the case of cosmological large-scale field
(smaller number) and knowledge on stronger fields associated
with the large-scale galaxy structure (larger
number). Eq.~(\ref{tsscale}) constitutes an interesting range of
emission timescales for many conceivable scenarios of UHECRs.
For example, the hot spots in certain
powerful radio galaxies that have been suggested as UHECR
sources~\cite{rb}, have a size of only several
kpc and could have an episodic activity on timescales of
$\sim10^6\,$yr.

A detailed comparison of analytical estimates for the distributions
of time delays, energies, and deflection angles of nucleons in
weak random magnetic fields with the results of Monte Carlo
simulations has been presented in Ref.~\cite{agnm}. In this
work, deflection was simulated by solving a stochastic differential
equation and observational consequences for the two major
classes of source scenarios, namely continuous and impulsive
UHECR production, were discussed. In agreement with earlier
work~\cite{mw} it was pointed out that at least in the impulsive
production scenario and for an EGMF in the range
$0.1-1\times10^{-9}\,$G, as required for cosmological GRB sources,
there is a typical energy scale
$E_b\sim10^{20.5}-10^{21.5}\,$eV below which the flux is
quasi-steady due to the spread in arrival times, whereas above
which the flux is intermittent with only a few sources
contributing.

\subsection*{General Case}

Unfortunately, neither the diffusive limit nor the limit of
nearly rectilinear propagation is likely to be applicable
to the propagation of UHECRs around $10^{20}\,$eV in general.
This is because in magnetic fields in the range of a few
$10^{-8}\,$G, values that are realistic for the Supergalactic
Plane~\cite{rkb,bbo}, the gyro radii of charged particles
is of the order of a few Mpc which is comparable to the distance to
the sources. An accurate, reliable treatment in this regime
can only be achieved by numerical simulation.

To this end,
the Monte Carlo simulation approach of individual trajectories
developed in Refs.~\cite{slo,sl} has recently been generalized
to arbitrary deflections~\cite{slb}. The Supergalactic Plane
was modeled as a sheet with a thickness of a few Mpc and a
Gaussian density profile. The same statistical description
for the magnetic field was adopted as in Refs.~\cite{slo,sl},
but with a field power law index $n_H=-11/3$, representing a
turbulent Kolmogorov type spectrum, and weighted with the
sheet density profile. It should be mentioned, however, that
other spectra, such as the Kraichnan spectrum,
corresponding to $n_H=-7/2$, are also possible. The largest mode
with non-zero power was taken to be the largest turbulent eddy
whose size is roughly the sheet thickness. In addition, a coherent
field component $B_c$ is allowed that is parallel to the sheet
and varies proportional to the density profile.

\begin{figure}[htb]
\centerline{\hbox{\psfig{figure=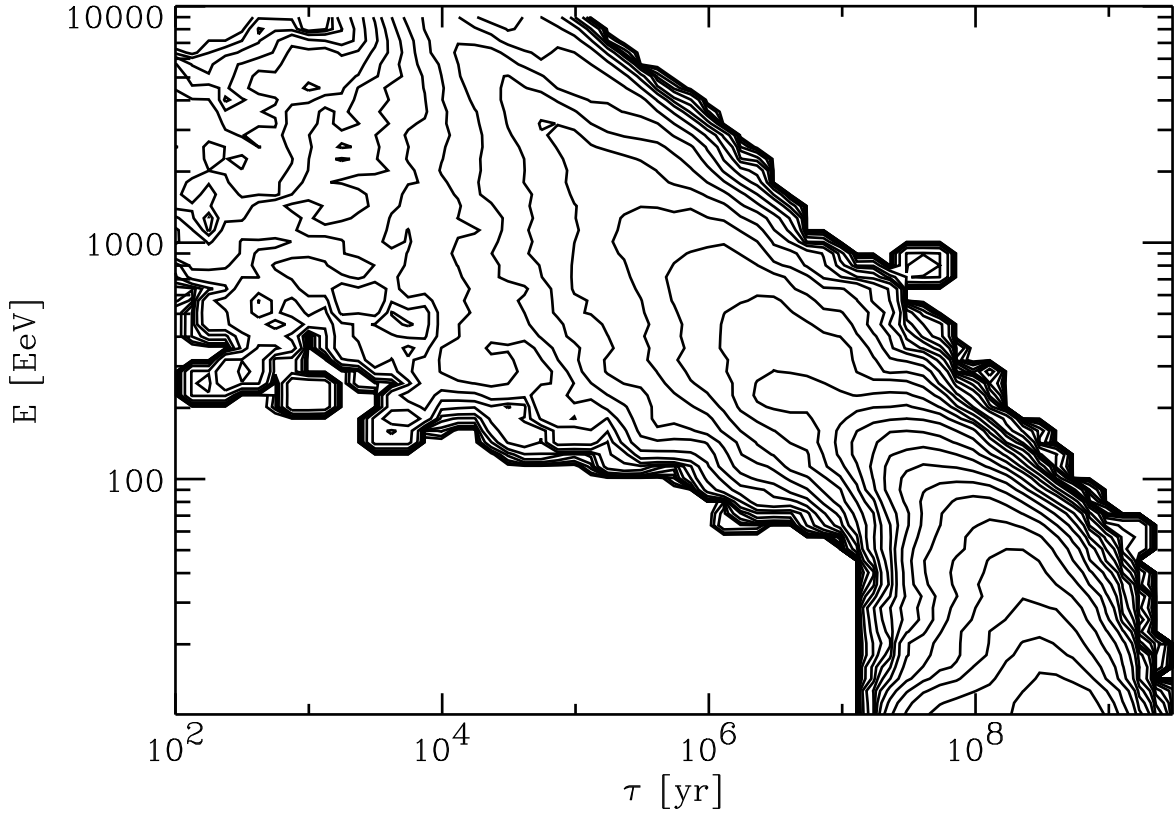,height=0.5\textheight}}}
\caption[...]{\footnotesize The distribution of time delays $\tau_E$
and energies $E$ for a burst with spectral index $\gamma=2.4$ at a distance
$d=10\,$Mpc, similar to Fig.~\ref{fig7}, but for the Supergalactic
Plane scenario discussed in the text. The turbulent magnetic field component
in the sheet center is $B=3\times10^{-7}\,$G. Furthermore,
a vanishing coherent field component is assumed.
The inter-contour interval is 0.25 in the logarithm to base 10 of the
distribution per logarithmic energy and time interval.
The three regimes discussed in the text,
$\tau_E\propto E^{-2}$ in the rectilinear regime $E\ga200\,$EeV,
$\tau_E\propto E^{-1}$ in the Bohm diffusion regime $60\,{\rm EeV}
\la E\la200\,$EeV, and $\tau_E\propto E^{-1/3}$ for
$E\la60\,$EeV are clearly visible.}
\label{fig11}
\end{figure}

When CR backreaction on the weakly turbulent magnetic field
is neglected, the diffusion coefficient of CR of energy $E$
is determined by the magnetic field power on wavelengths
comparable to the particle Larmor radius, and can be approximated
by
\begin{equation}
  D(E)\simeq\frac{1}{3}\,r_g(E)\,
  \frac{B}{\int_{1/r_g(E)}^\infty
  \,dk\,k^2\left\langle B^2(k)\right\rangle}\,.\label{Danaly}
\end{equation}
As a consequence, for the Kolmogorov spectrum, in the diffusive
regime, where $\tau_E\ga d$, the diffusion coefficient
should scale with energy as $D(E)\propto E^{1/3}$ for
$r_g\la L/(2\pi)$, and as $D(E)\propto E$ in the so called
Bohm diffusion regime,$r_g\ga L/(2\pi)$. This should be
reflected in the dependence of the time delay $\tau_E$ on
energy $E$: From the rectilinear regime, $\tau_E\la d$,
hence at the largest energies, where $\tau_E\propto E^{-2}$,
this should switch to $\tau_E\propto E^{-1}$ in the regime of
Bohm diffusion, and eventually to $\tau_E\propto E^{-1/3}$
at the smallest energies, or largest time delays. Indeed,
all three regimes can be seen in  Fig.~\ref{fig11} which shows
an example of the distribution of arrival times and energies
of UHECRs from a bursting source.

The numerical results indicate an effective
gyroradius that is roughly a factor 10 higher than the
analytical estimate, with a correspondingly larger diffusion
coefficient compared to Eq.~(\ref{Danaly}). In addition,
the fluctuations of the resulting spectra 
between different magnetic field realizations
can be substantial. This is a result of the fact that
most of the magnetic field power is on the largest scales
where there are the fewest modes. These considerations mean that 
the applicability of analytical flux estimates of discrete sources in
specific magnetic field configurations is rather limited. 

\begin{figure}[htb]
\centerline{\hbox{\psfig{figure=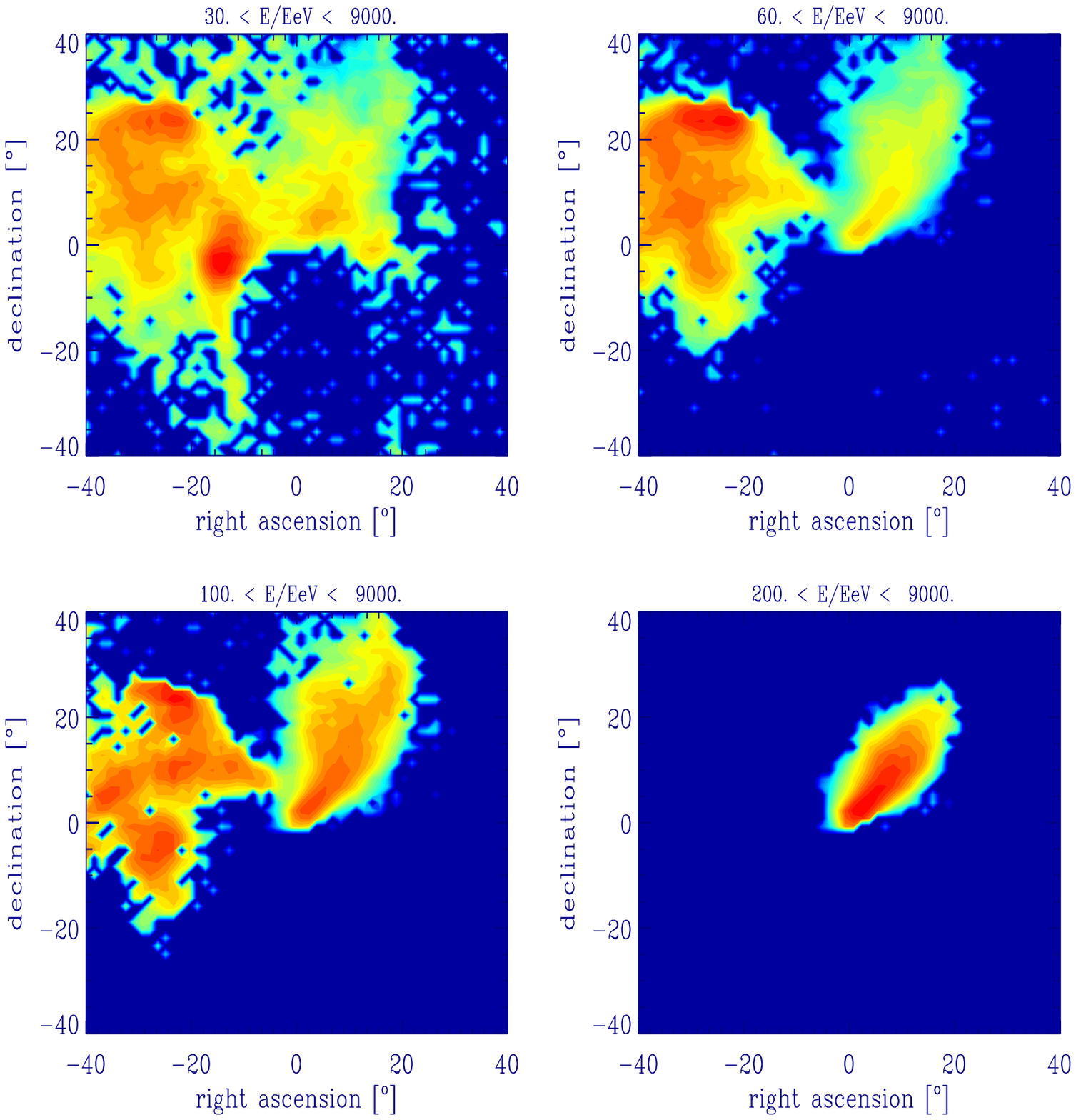,height=0.75\textheight}}}
\caption[...]{\footnotesize Angular image of a point-like source in
a magnetized Supergalactic Plane, corresponding to
one particular magnetic field realization with
a maximal magnetic field in the plane center, 
$B_{\rm max}=5\times10^{-8}\,$G,
all other parameters being the same as in Fig.~\ref{fig11}.
The image is shown in different energy ranges, as indicated,
as seen by a detector of $\simeq1^\circ$ angular resolution.
A transition from several images at lower energies to only one image at the
highest energies occurs where the linear deflection becomes
comparable to the effective field coherence length.
The difference between neighboring shade levels is 0.1 in the
logarithm to base 10 of the integral flux per solid angle.}
\label{fig12}
\end{figure}

In a steady state situation, diffusion leads to a modification
of the injection spectrum by roughly a factor $\tau_E$, at
least in the absence of significant energy loss and for a
homogeneous, infinitely extended medium that can be described
by a spatially constant diffusion coefficient. Since in the
non-diffusive regime the observed spectrum repeats the shape of
the injection spectrum,
a change to a flatter observed spectrum at high energies is
expected in the transition region~\cite{diff_refs}. From
the spectral point of view this
suggests the possibility of explaining the observed UHECR
flux above $\simeq10\,$EeV including the highest energy
events with only one discrete source~\cite{bo}.

Angular images of discrete sources in a magnetized Supercluster
in principle contain information on the magnetic field structure.
For the recently suggested field strengths between
$\sim10^{-8}\,$G and $\simeq1\mu\,$G the angular images are
large enough to exploit that information with instruments
of angular resolution in the degree range. An example where
a transition from several images at low energies to one image
at high energies allows one to estimate the magnetic field
coherence scale is shown in Fig.~\ref{fig12}.

The newest AGASA data~\cite{agasa2}, however, indicate an
isotropic distribution of EHECR. To explain this with only
one discrete source would require the magnetic fields to be
so strong that the flux beyond $10^{20}\,$eV would most likely
be too strongly suppressed by pion production, as discussed above.
The recent claim that the powerful radio galaxy Centaurus A
at a distance of 3.4 Mpc from Earth can explain both observed
flux and angular distributions of UHECRs above $10^{18.7}\,$eV
for an extragalactic magnetic field of strength
$\simeq0.3\mu\,$G~\cite{fp}
was based on the diffusive approximation which is clearly
shown by numerical simulations not to apply in this situation;
the predicted distributions are too anisotropic even at
$10^{19}\,$eV.
  
This suggests a more continuous source distribution which
may also still reproduce the observed UHECR
flux above $\simeq10^{19}\,$eV with only one spectral
component~\cite{medinatanco3}. Statistical studies neglecting
magnetic deflection can be performed analytically and
also suggest that many sources should contribute to the
UHECR spectrum~\cite{stat_analy}. A more systematic parameter study
of sky maps and spectra in UHECR in different scenarios with
magnetic fields is now being pursued~\cite{medinatanco4,lsb}.
Such studies will also be needed to generalize analytical
considerations on correlations of UHECR arrival directions
with the large scale structure of galaxies~\cite{lss_corr}
to scenarios involving significant magnetic deflection.

Intriguingly, scenarios in which a diffuse
source distribution follows the density in the Supergalactic
Plane within a certain radius, can accomodate both the large scale
isotropy (by diffusion) and the small scale clustering (by magnetic
lensing) revealed by AGASA if
a magnetic field of strength $B\ga0.5\mu\,$G
permeates the Supercluster~\cite{lsb}.

\begin{figure}[ht]
\centerline{\hbox{\psfig{figure=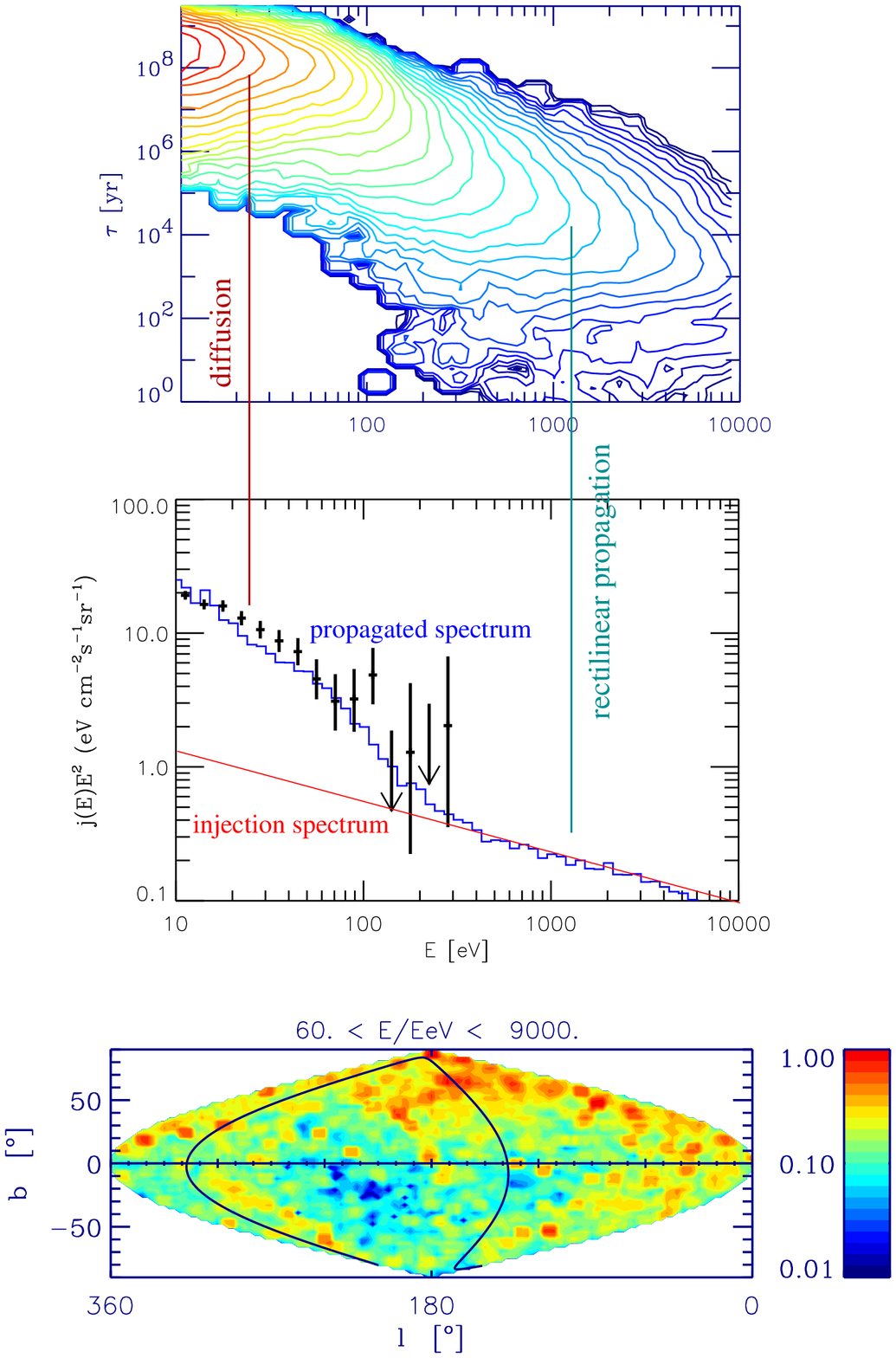,height=0.85\textheight}}}
\caption[...]{\footnotesize The distribution of arrival times and
energies (top), the solid angle integrated spectrum (middle,
with 1 sigma error bars showing combined data
from the Haverah Park~\cite{haverah}, the Fly's Eye~\cite{fe},
and the AGASA~\cite{agasa} experiments above $10^{19}\,$eV),
and the angular distribution of
arrival directions in Galactic coordinates (bottom,
with color scale showing the intensity per solid angle)
in the Supercluster scenario with continuous source
distribution explained in the text,
averaged over 4 magnetic field realizations
with 20000 particles each.}
\label{fig13}
\end{figure}

Fig.~\ref{fig13} shows the distribution of arrival times and energies,
the solid angle integrated spectrum, and the angular distribution of
arrival directions in Galactic coordinates in such a scenario
where the UHECR sources with spectral index $\gamma=2.4$ are
distributed according to the matter density in the Local
Supercluster, following a pancake profile with scale
height of 5 Mpc and scale length 20 Mpc.
The r.m.s. magnetic field has a Kolmogorov spectrum with a maximal field
strength $B_{\rm max}=5\times10^{-7}\,$G in the plane center,
and also follows the matter density.
The observer is within 2 Mpc of the Supergalactic Plane whose location is
indicated by the solid line in the lower panel and at a distance
$d=20\,$Mpc from the
plane center. The absence of sources within 2 Mpc from the observer
was assumed. The transition discussed above from the diffusive
regime below $\simeq2\times10^{20}\,$eV
to the regime of almost rectilinear propagation above that energy
is clearly visible.

Detailed Monte Carlo simulations performed on these distributions
reveal that the anisotropy decreases with increasing magnetic field
strength due to diffusion and that small scale clustering increases
with coherence and strength of the magnetic field due to magnetic
lensing. Both anisotropy and clustering also increase with
the (unknown) source distribution radius.
Furthermore, the discriminatory power
between models with respect to anisotropy and clustering strongly
increases with exposure~\cite{lsb}.

As a result, a diffuse source distribution associated with the
Supergalactic Plane can explain most of the currently observed 
features of ultra-high
energy cosmic rays at least for field strengths close to $0.5\,\mu\,$G.
The large-scale anisotropy and the clustering predicted
by this scenario will allow strong discrimination against other models
with next generation experiments such as the Pierre Auger Project.

\clearpage

\clearpage

\section{Conclusions}

Ultra-high energy cosmic rays have the potential to open a window
to and act as probes of new particle physics beyond the Standard Model
as well as processes occurring in the early Universe at energies close
to the Grand Unification scale. Even if their origin will turn out
to be attributable to astrophysical shock acceleration with no new
physics involved, they will still be witnesses of one of the most energetic
processes in the Universe. Furthermore, complementary to other methods
such as Faraday rotation measurements, ultra-high energy cosmic
rays can be used as probes of the poorly known large
scale cosmic magnetic fields. The future appears promising and
exciting due to the anticipated arrival of several large scale
experiments.

\end{document}